\documentclass{emulateapj}
\pdfoutput=1

\usepackage[pdftex,bookmarks=true]{hyperref}
\usepackage{natbib} 

\begin{document}

\renewcommand{\arraystretch}{1.2}

\title{QSO Selection Algorithm Using Time Variability and Machine Learning:\\
Selection of 1,620 QSO Candidates from MACHO LMC Database}

\author{Dae-Won Kim\altaffilmark{1}\altaffilmark{,2}\altaffilmark{,3},
Pavlos Protopapas\altaffilmark{1}\altaffilmark{,3},
Yong-Ik Byun\altaffilmark{2},
Charles Alcock\altaffilmark{1},
Roni Khardon\altaffilmark{4},
Markos Trichas\altaffilmark{1}}
\affil{\altaffilmark{1}Harvard-Smithsonian Center for Astrophysics, Cambridge, MA, USA}
\affil{\altaffilmark{2}Department of Astronomy, Yonsei University, Seoul, South Korea}
\affil{\altaffilmark{3}Institute for Applied Computational Science, Harvard University, Cambridge, MA, USA}
\affil{\altaffilmark{4}Department of Computer Science, Tufts University, Medford, MA, USA}

\begin{abstract}

We present a new QSO  selection algorithm using a Support Vector Machine (SVM), 
a supervised classification method, on a set of extracted time series features  
including period, amplitude, color, and autocorrelation value.
We  train a model that separates QSOs from variable stars,  non-variable stars
and microlensing events using 58 known QSOs,  
1,629 variable stars and 4,288 non-variables using the  
MAssive Compact Halo Object (MACHO) database as a training set.
To estimate the efficiency and the accuracy of the model, 
we perform a cross-validation test using the training set.
The test shows that the model correctly identifies $\sim$80\% 
of known QSOs with a 25\% false positive rate.
The majority of the false positives are Be stars.

We applied the trained model to the MACHO Large Magellanic Cloud (LMC) dataset, 
which consists of 40 million lightcurves, and found 1,620 QSO candidates.
During the selection none of the 33,242 known 
MACHO variables were misclassified as QSO candidates.
In order to estimate the true false positive rate, 
we crossmatched the candidates with astronomical catalogs including 
the Spitzer Surveying the Agents of a 
Galaxy's Evolution (SAGE) LMC catalog and a few X-ray catalogs. 
The results further suggest that the majority of the candidates, more than 70\%, are QSOs.

\end{abstract}

\keywords{Magellanic Clouds - methods: data analysis - quasars: general}

\section{Introduction}

A large catalog of Quasi-stellar object (QSO)
is important for a variety of fields in modern astrophysics and observational cosmology.
QSOs have been used for studies of
a) large scale structures based on the spatial clustering of QSOs
\citep{Shen2007AJ, Ross2009ApJ},
b) growth of central black holes using the estimated black holes' masses
\citep{Kollmeier2006ApJ},
c) coevolution of black holes and their host galaxies using lensed QSO hosts
\citep{Peng2006ApJ},
d) the epoch of reionization based on high redshift QSOs
\citep{Becker2001AJ, Fan2006AJ},
e) dark matter substructure using gravitationally lensed QSOs
\citep{Metcalf2001ApJ, Miranda2007MNRAS} and 
f) properties of the intergalactic medium determined by measuring metallicity distribution using QSO spectra
\citep{Viel2002MNRAS, Simcoe2004ApJ}.

One of the most interesting properties of QSOs is 
the strong flux variation over a wide range of 
wavelengths on timescales from days to years 
(\citealt{Hook1994MNRAS, Hawkins2002MNRAS} and references therein). 
It is believed that QSO variability is associated with accretion disk instabilities
\citep{Rees1984ARAA, Kawaguchi1998ApJ}
although there are other possible explanations for the source of QSO variability, 
including microlensing
\citep{Hawkins1993Nature, Zackrisson2003AA},
starbursts and supernovae 
\citep{Terlevich1992MNRAS, Aretxaga1997MNRAS}.
It is debatable which mechanism is the dominant source of variability
(see \citealt{Hook1994MNRAS, Giveon1999MNRAS, VandenBerk2004ApJ, DeVries2005AJ, Bauer2009ApJ}).
Moreover, due to the lack of  long-time-span, well-sampled and high-quality QSO lightcurves, 
all these previous studies have investigated ensemble variabilities of QSOs.
Thus it is important to have a large set of well-sampled 
QSO lightcurves in order to study both ensemble and individual QSO variability characteristics,
which will help constrain the theoretical models of the variability mechanisms
(see \citealt{Hook1994MNRAS, Cristiani1996AA, VandenBerk2004ApJ} and references therein).

Many authors have attempted to select QSO candidates based on the variability characteristics.
For instance, \citet{Eyer2002AcA} selected QSO candidates from 
68,000 OGLE-II variable stars \citep{Zebrun2001AcA}
using colors, magnitudes and the structure function of the variables.
The structure function determines the time scale
of variability in a given lightcurve as a function of the time lag 
between observations \citep{Eyer2002AcA}.
Among the selected 133 QSO candidates, $\sim$10\% 
were confirmed to be QSOs \citep{Dobrzycki2002ApJ,Dobrzycki2005AA}.
\citet{Geha2003AJ} (hereinafter G03) searched 140,000 MACHO sources that have
significant flux variation \citep{Alcock2000ApJ}.
G03 used colors, magnitudes and two statistical parameters 
that quantify variability to select QSO candidates.
G03 then removed known MACHO variable stars \citep{Alcock2001} from the candidate list 
and finally examined the remaining  candidates manually in order to remove false positives. 
Among the final 360 candidates, 259 were spectroscopically observed and
47 of them confirmed to be QSOs.
\citet{Sumi2005MNRAS} searched about 200,000 variable objects of the OGLE-II data \citep{Wozniak2002AcA}
and then used a few selection cuts such as magnitudes, structure function and manual validation.
No spectroscopic observation was done for their final 97 QSO candidates.

Recently, four QSO selection methods have been submitted or published,
which proposed new QSO classification algorithms using time series variability features.
One of them is the work done by \citet{Kozlowski2010ApJ} that used a stochastic model
shown in \citet{Kelly2009ApJ} which derives the amplitude and the time scale of lightcurve variations.
They also employed periods  of lightcurves and magnitudes.
To develop their selection method, they used the known QSOs,
periodic variables and non-periodic variables in the OGLE databases 
\citep{Udalski1997AcA, Udalski2008AcA}.
They also used QSO candidates from \citet{Kozlowski2009ApJ} that had OGLE counterparts.
To separate the QSOs from other variables, 
they defined several cuts and correctly identified 63\% of 
the QSOs while removing most of the variable stars.
The second study \citep{Schmidt2010ApJ} proposed a power-law model to fit the structure function
and derived the amplitude and the power index of the model.
They used the derived parameters to isolate 
known QSOs from RR Lyraes and non-variable stars 
extracted from the SDSS stripe 82 database (S82) \citep{Sesar2007AJ}.
Using simple cuts on the amplitude versus power index plane, 
they identified about 90\% of the SDSS QSOs with a 5\% false positive rate.
 \citet{Butler2010} and \citet{MacLeod2010}
used  similar approaches (i.e. structure function) with the previous two works.
Both utilized the preselected variable sources from the S82 dataset 
where the majority of the variables are
QSOs, RR Lyraes and stars from the stellar locus (see \citealt{Sesar2007AJ} for details).
\citet{Butler2010} parameterized the ensemble QSO structure 
function as a function of brightness of the QSOs.
They then used the parameterized ensemble QSO model 
to evaluate the quasar likelihood for individual lightcurves
(see \citealt{Butler2010} for details).
Using this method, they identified nearly all the known SDSS QSOs (99\%) with a 3\% false positive rate.
\citet{MacLeod2010} also used the structure function
and several cuts to identify QSOs and exclude other variable stars from the S82 database.
They correctly selected about 90\% of the QSOs 
with 10$\sim$20\% false positive rate depending on the cuts imposed.
Both works also selected new QSO candidates from the preselected variable sources \citep{Sesar2007AJ}.
These candidates have not been spectroscopically confirmed.
Note that the efficiencies or false positive rates of these studies
should not be directly compared because each work used 
their own selected set of stars and QSOs to develop their methods.
For a comprehensive comparison of the results of the methods, see \citet{MacLeod2010}.

Even though some of these recent works \citep{Schmidt2010ApJ, Butler2010, MacLeod2010} showed
high efficiencies and low false positive rates, 
they used samples that are selected in such a way 
that high efficiency and low false positive rate is to be expected.
The separation of QSOs from non-varying stars and a few types of variable stars, 
especially short-period variables (i.e. RR Lyraes) 
are relatively straightforward since QSOs show non-periodic and long-time scale fluctuation.
The majority of the samples they used in these studies are short-period variables
and do not show long-time scale fluctuation,

QSO selection methods based on variability will be valuable tools
for on-going and future large scale survey missions
such as Pan-STARRS \citep{Kaiser2004SPIE} and LSST \citep{Ivezic2008arXiv}.
These surveys will keep monitoring wide areas of the sky 
and will produce vast amount of time series data 
in several wavelength bands (e.g. $g, r, i, z$ for Pan-STARRS).
Because spectroscopic observations for such wide areas are very expensive, 
QSO selections in the absence of spectroscopic data are becoming important,
and thus developing QSO selection methods using variability
are rapidly attracting notable attention.

The work presented in this paper utilizes the whole MACHO lightcurve databset considering all known
variable sources in the MACHO database.
Thus this is the first work that considers the efficiency and the false positive rates 
of QSO
selection in an entire lightcurve dataset.
We have developed our method by training on  the richest possible dataset
including all known types of sources
and testing it also on the whole dataset.
The training set includes a variety of variable objects such as
QSOs, RR Lyraes, Cepheids, eclipsing binaries, long period variables, Be stars, 
microlensing events and also non-variable stars.
Only one other selection method, \citet{Kozlowski2010ApJ},
has considered Be stars, which are
one of the most significant contaminants during QSO selections in LMC (G03).
Our goal is to select high confidence QSO candidates
in the MACHO database \citep{Alcock1996ApJ}
while minimizing the number of false positives.
Our approach employs multiple time series features rather than
using only the lightcurve structure function.
These features can characterize various kinds of variability characteristics.
Therefore our algorithm is practical not only for identifying QSOs
but also for excluding other types of variable stars and non-variable stars. 
To fully utilize the features and identify QSOs,
we employed a supervised machine learning classification method, 
Support Vector Machine (SVM, \citealt{Boser1992, Cristianinic2000, Panik2005}).
In the true spirit of machine learning, our method 
uses a classification model trained with the training set
and thus eliminates the need for hard linear cuts 
and  human input (e.g.  manual preselection of variable sets, 
manual removal of false positives and determination of cuts).

We briefly introduce the MACHO database and known MACHO QSOs in Section \ref{sec:MACHO}.
Section \ref{sec:Features} describes the multiple time series features 
that we used to quantify the variability characteristics of each lightcurve.
Section \ref{sec:Support_Vector_Machine} introduces SVM,
the method used to train the classification model.
We present the MACHO QSO classification model constructed using the time series features and SVM in Section \ref{sec:MACHO_QSO_SVM_Models}.
We then show the MACHO QSO candidates selected using the model in Section \ref{sec:MACHO_QSO_Candidates_Selection}.
Crossmatched results with astronomical catalogs are presented in Section \ref{sec:crossmatching}.
Ongoing and future work is summarized in Section \ref{sec:Future_Works}.

\begin{table*}
\begin{center}
\caption{11 time series features\label{tab:features}}
\begin{tabular}{cl}
\tableline\tableline
Four new features & Brief description; for details, see Appendix. \\
\tableline
$N_{above}$ and $N_{below}$ & $N_{above}$: the number of points above the upper boundary line of the autocorrelation plot. \\
	& $N_{below}$: the number of points below the lower boundary line of the autocorrelation plot. \\
	& Figure \ref{fig:AC_boundary} shows the constructed boundary lines based on the autocorrelation functions (see Figure \ref{fig:AC}) \\
	& of the training set lightcurves. \\\\

Stetson $K_{AC}$ & Stetson $K$ (Eq. \ref{eq:StetsonK}) variability index derived based on the autocorrelation function \\
	&of each lightcurve. \\\\

$R_{cs}$ & The range of a cumulative sum \citep{Ellaway1978}. \\\\

\tableline\tableline
Seven other features & Brief description; for details, see Appendix. \\
\tableline 

$\sigma / \bar{m}$ & The ratio of the standard deviation, $\sigma$, to the mean magnitude, $\bar{m}$. \\\\

Period and Period S/N & Period and period signal-to-noise ratio of each lightcurve.\\
	& Derived using Lomb-Scargle algorithm and Lomb periodogram 
	\citep{Lomb1976ApSS, Scargle1982ApJ}. \\\\

Stetson $L$ &  The variability index \citep{Stetson1996PASP} describes the synchronous variability of different bands. \\\\

$\eta$ & The ratio of the mean of the square of successive differences to the variance of data points \\
	& in each lightcurve. \\\\

$B-R$ & The average color for each lightcurve. \\\\ 

$Con$ & The number of  consecutive data points that are brighter or fainter than 2$\sigma$ of each lightcurve. \\

\tableline
\end{tabular}
\end{center}
\end{table*}

\section{MACHO database and MACHO QSO}
\label{sec:MACHO}

\subsection{MACHO database}

The MACHO survey monitored a wide area of the sky to detect microlensing events
caused by Milky Way halo objects and test the hypothesis that 
a significant portion of dark matter in the Milky Way halo consists of compact objects
such as brown dwarfs or planets \citep{Alcock1996ApJ}. Because microlensing events are extremely
rare, MACHO monitored several tens of millions of stars in the Large Magellanic Cloud (LMC), 
Small Magellanic Cloud (SMC) and Galactic bulge for 7.4 years.
Observations started in July 1992 and were completed at the end of December 1999.
More than 5 Tbytes of image data and 70,000 exposures were collected during the period \citep{Alcock2000ApJ}.
In addition, MACHO used two bands (MACHO B and R) for the observations.

\subsection{{\rm{MACHO QSOs}}}

There are in total 59 known QSOs in the MACHO database (50 in the LMC fields and 9 in the SMC fields; hereinafter MACHO QSOs). 
Forty-seven  were detected by G03 and the remaining twelve were QSOs 
previously known from other studies \citep{Blanco1986PASP, Schmidtke1999AJ, Dobrzycki2002ApJ}. 
G03 detected 38 of them using variability characteristics of MACHO lightcurves and
nine of them by crossmatching with X-ray and radio catalogs.
To select  QSO candidates,
G03  applied simple cuts such as color, magnitude and amplitude
on 140,000 preselected MACHO sources  that show strong flux variation \citep{Alcock2000ApJ}.
The lightcurves of 12 previously known QSOs
were used as references for the variability cuts.
After selecting 2,500 QSO candidates from the 140,000 sources,
G03 removed known MACHO variable stars from the candidate list
and then manually examined the remaining candidates to eliminate false positives.
They eventually removed about 2,140 candidates and 
confirmed that the majority of the removed candidates were objects with quasi-periodic
variability such as blue variable stars.
Blue variable stars typically show strong Balmer emission lines
and are thought to be associated with Be stars \citep{Keller2002AJ}.
It is also known that Be stars show variability similar to QSOs 
\citep{Eyer2002AcA, Geha2003AJ, Mennickent2002AA, Keller2002AJ}.
Using spectroscopic instruments, G03 observed 259 candidates selected from the 
remaining 360 candidates and also the candidates selected using the catalog crossmatchings.
G03 confirmed 47 new QSOs with magnitudes $16.63 < m_{V} < 20.10$
and redshifts between 0.28 and 2.77.

G03 analyzed only 30 of the 82 MACHO LMC fields,
and thus the remaining 52 MACHO LMC fields have not been searched for QSOs.
Moreover, they selected QSO candidates from the preselected 140,000 variable sources
and did not analyze the remaining several tens of million lightcurves. 
Thus it is very likely that there are a lot more QSOs that have not been detected yet.
In the following sections, we introduce a new QSO selection algorithm
to detect these non-identified QSOs in the MACHO LMC database.

\section{Time Series Features}
\label{sec:Selection_Algorithm}

In order to separate QSOs from non-variable stars and variable stars,
we quantify the variability characteristics of lightcurves using 11 time series features.
These 11 features were independently proposed to quantify certain types of variability features
including amplitudes, periods, colors and distribution of data points.
They can complement each other because they pick out different variability features.
Thus, by using these multiple features, we can identify various types of variability characteristics
(e.g. non-varying sources, periodic variables and non-periodic variables).
Note that we selected these time series features not only for characterizing QSO time series
but also for characterizing other types of variable sources or non-variable sources
because we want to identify QSOs while excluding the other types of sources at the same time.
We briefly describe these 11 time series features in Table \ref{tab:features}.
See Appendix for details about the features 
consisting of four new features that we have developed for this work
and seven previously used features.

\begin{figure*} 
\begin{center}
\begin{minipage}[c]{12cm}
        \includegraphics[width=1.0\textwidth]{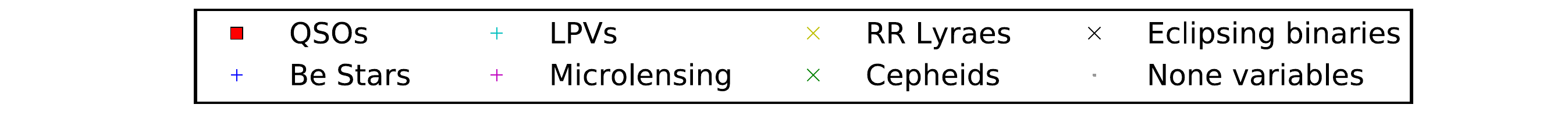} 
\end{minipage}
\\
\begin{minipage}[c]{8cm}
        \includegraphics[width=1.0\textwidth]{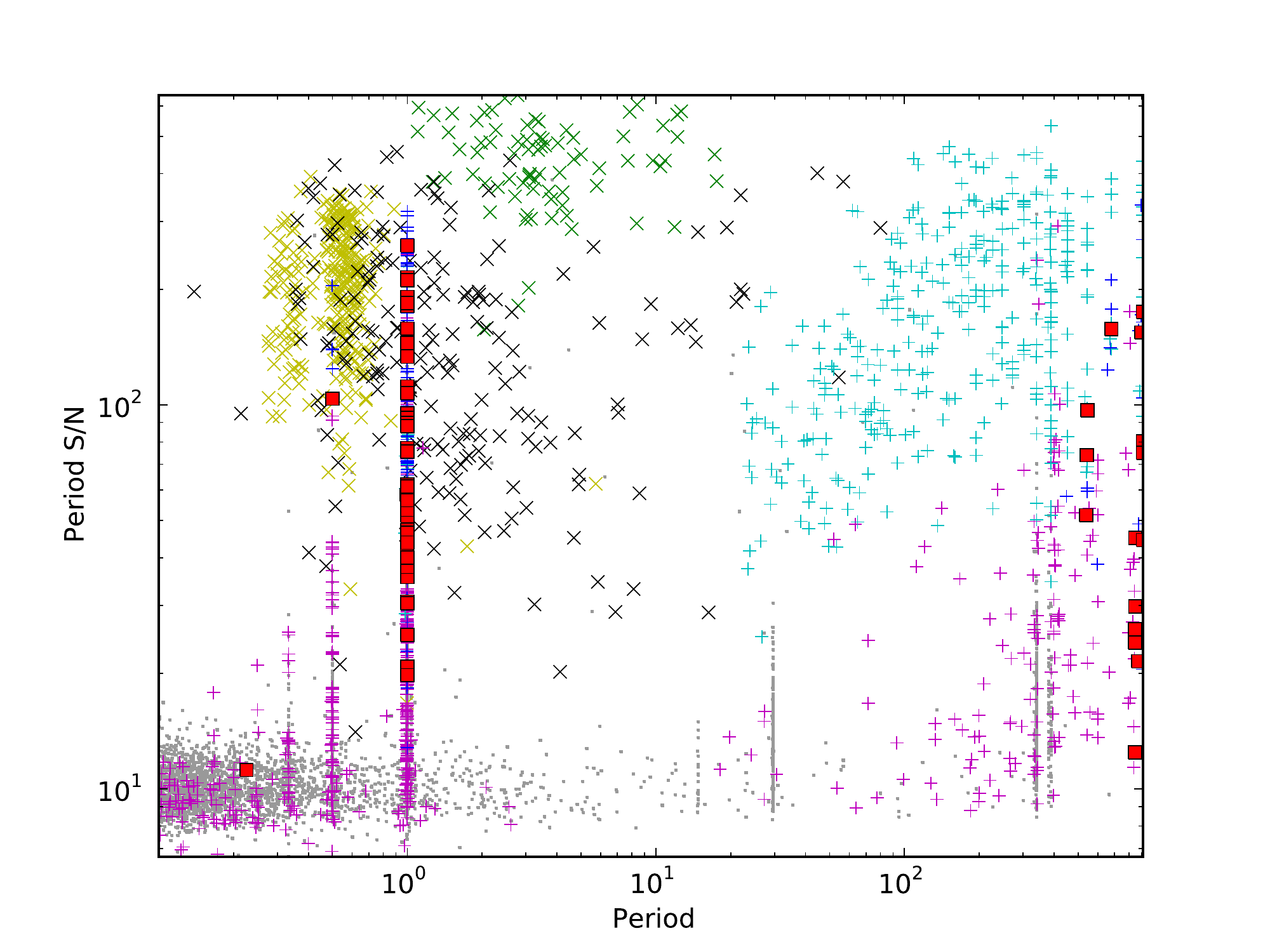}
\end{minipage}
\begin{minipage}[c]{8cm}
        \includegraphics[width=1.0\textwidth]{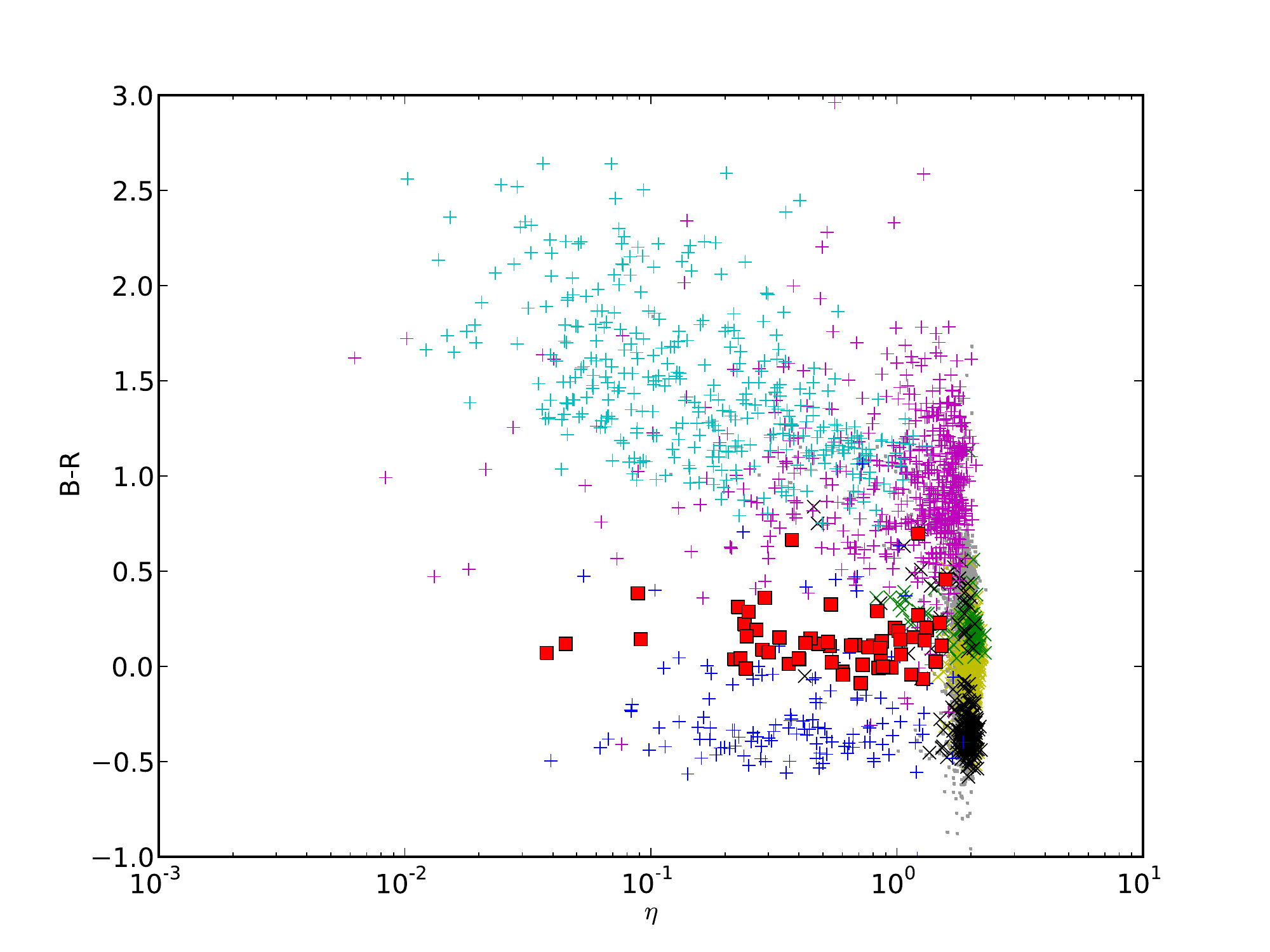}
\end{minipage}
\\
\begin{minipage}[c]{8cm}
        \includegraphics[width=1.0\textwidth]{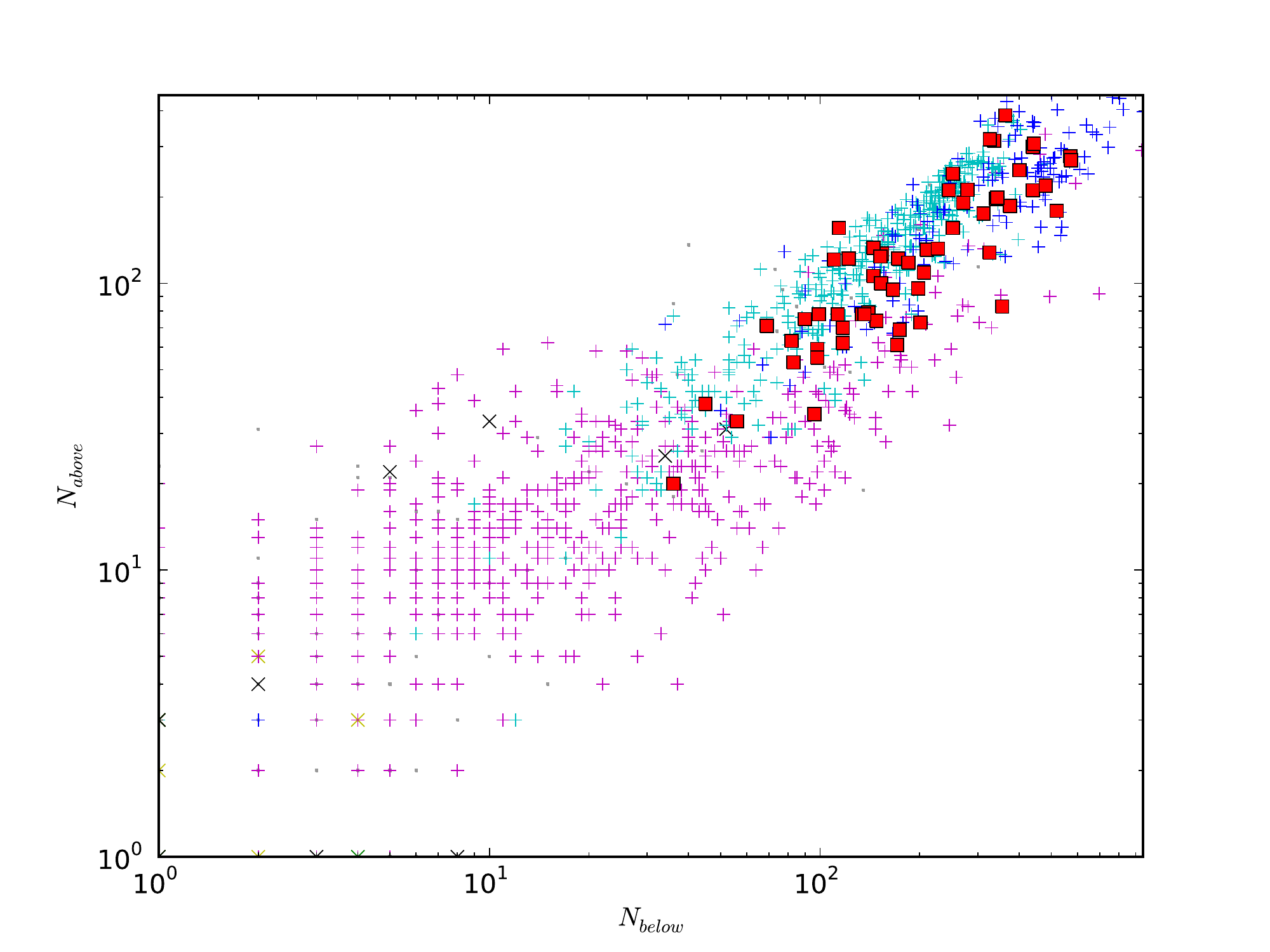}
\end{minipage}
\begin{minipage}[c]{8cm}
        \includegraphics[width=1.0\textwidth]{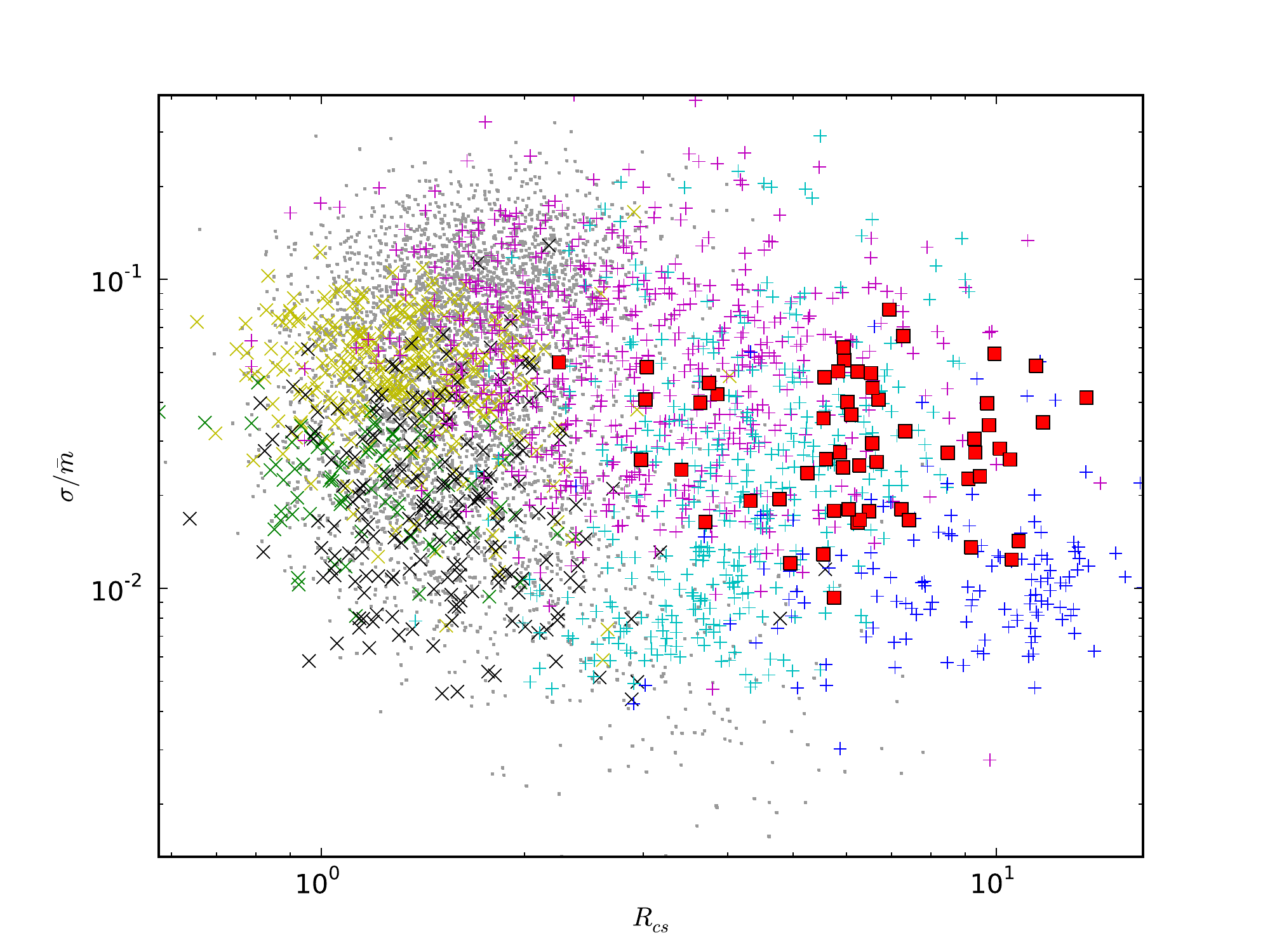}
\end{minipage}
\\
\begin{minipage}[c]{8cm}
        \includegraphics[width=1.0\textwidth]{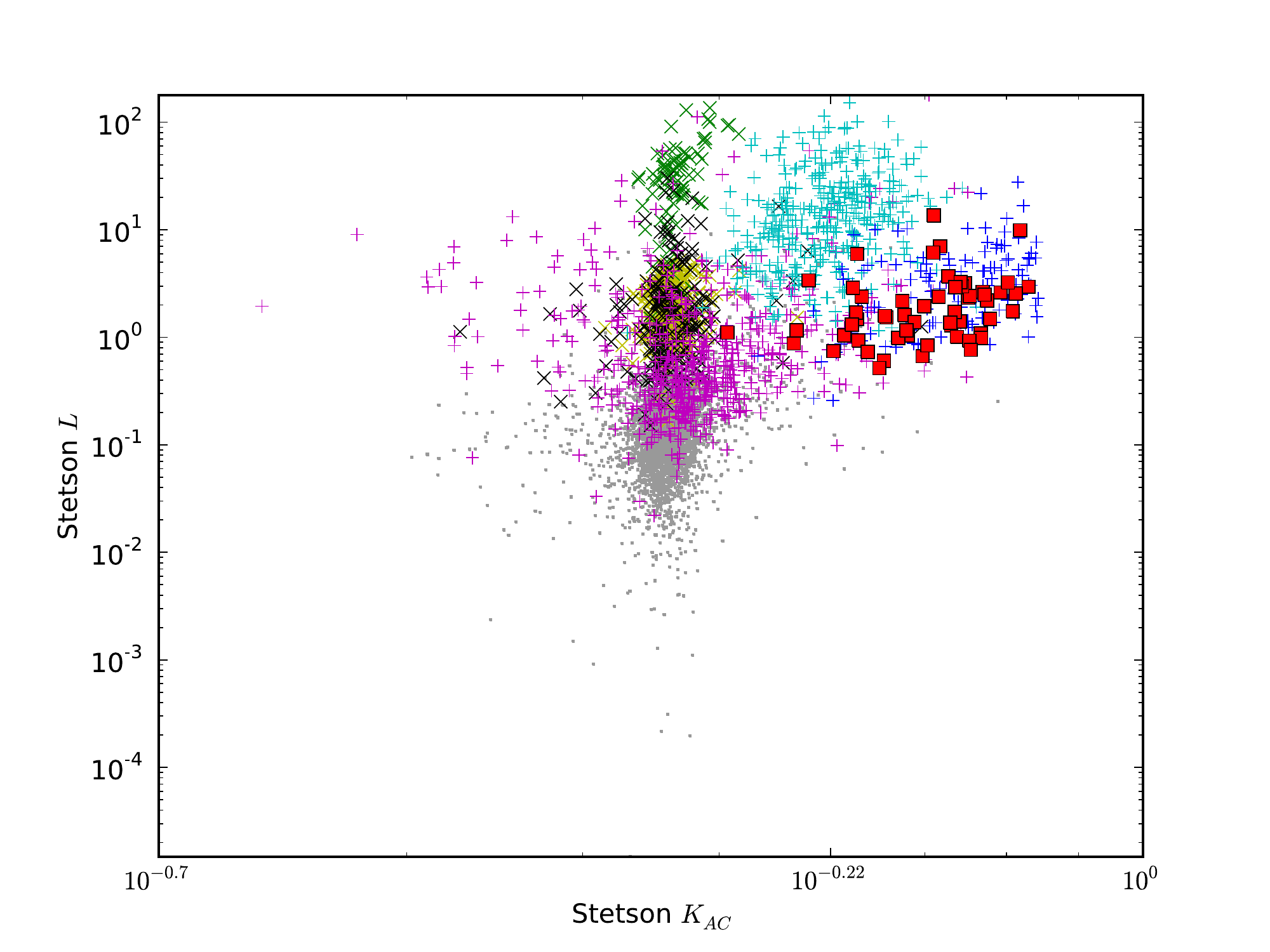}
\end{minipage}
\begin{minipage}[c]{8cm}
        \includegraphics[width=1.0\textwidth]{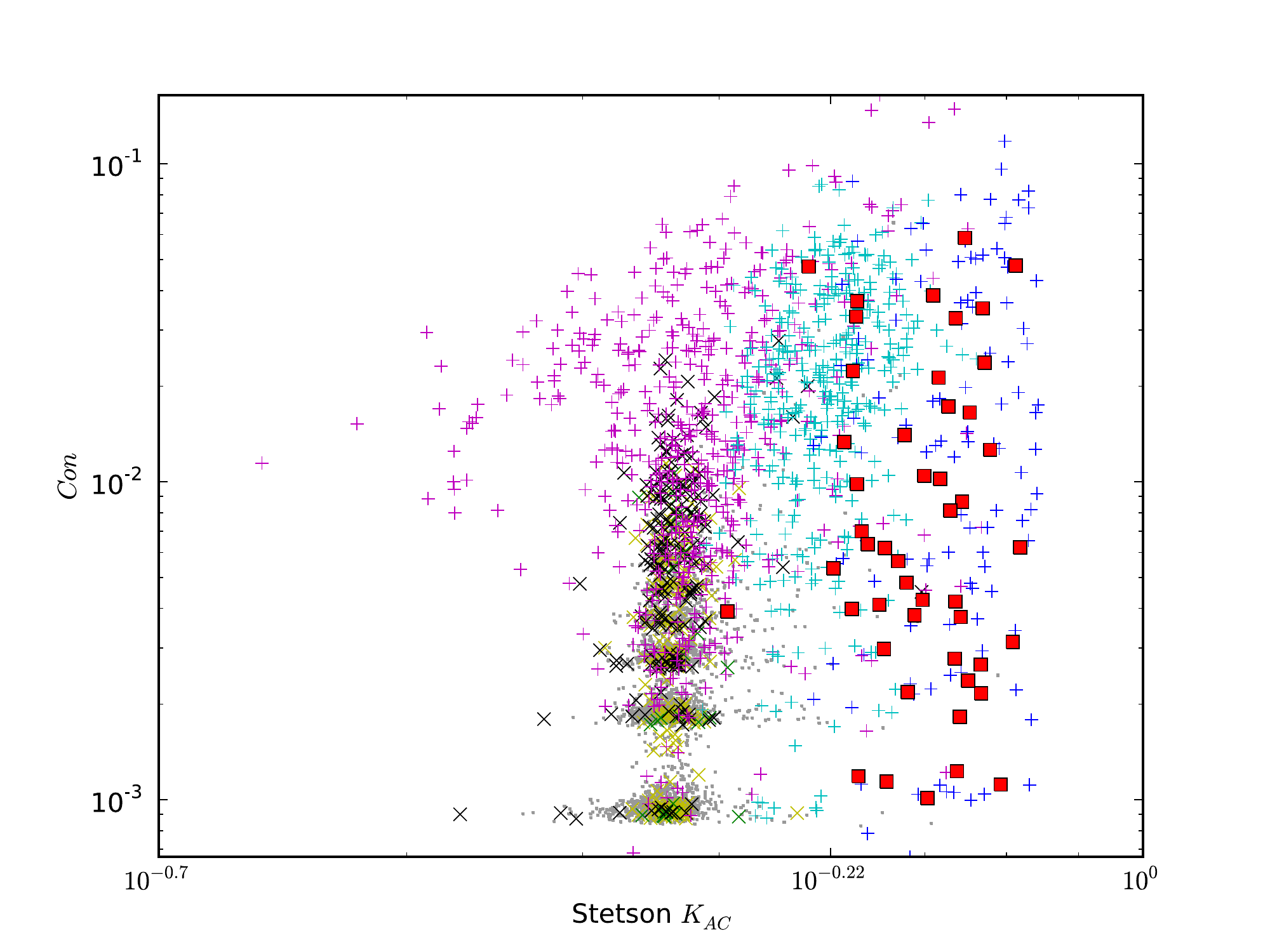}
\end{minipage}
\end{center}
    \caption
           {Scatter plots of the 11 time series features. The axis of each panel is different time series feature.
As the panels show, each type of variables is clustered in certain areas.
1) The top left panel: each type of the periodic variables are clustered at each different area. 
It also shows one day or multiple days period aliases caused by MACHO's observational nightly pattern.
2) The top right panel: $\eta$ is relatively small for QSOs, Be stars and LPVs
which have positive autocorrelation.
Color (i.e. difference between average magnitude of MACHO $B$ and $R$ bands) 
is useful in separating QSOs from some other types of variables as several other studies suggested 
\citep{Giveon1999MNRAS, Eyer2002AcA, Geha2003AJ}.
3) The middle left panel: $N_{above}$ vs $N_{below}$.
The panel shows almost none of the non-variables
and periodic variables except LPVs because they do not have data points above (below) 
the boundary lines  by construction. 
4) The middle right panel: $R_{cs}$ is relatively larger for QSOs.
$\sigma / \bar{m}$ also separates some variable types.
For instance, Be stars have relatively smaller values of $\sigma / \bar{m}$ than QSOs.
5) The bottom left panel: Stetson $L$ is effective in separating any type of variables from 
non-variables except microlensing events
while Stetson $K_{AC}$ is practical to separate QSOs, 
Be stars and LPVs from others.
6) The bottom right panel: $Con$ can be used  to separate non-variables from others
because non-variables have relatively smaller $Con$ than the others.
For details about each feature, see the text and Appendix.
}
    \label{fig:index_index}
\end{figure*}

Figure \ref{fig:index_index} shows scatter plots of all 11 time series features.
Different colors and symbols denote different types of sources.
The red squares are QSOs, the blue crosses are Be stars,
the magenta crosses are microlensing events, the cyan crosses are LPVs,
the green x's are Cepheids, the yellow x's are RR Lyraes, the black x's are eclipsing binaries
and the gray dots are non-variables.
As each panel shows, not only QSOs but also other types of variables are clustered in certain areas,
which means each time series feature is good at separating some of the variable types.
Thus we did not implement a feature selection algorithm that removes uninformative features. 
See Section \ref{sec:Support_Vector_Machine} for a brief explanation 
about a general feature selection concept.
We selected a subset of MACHO lightcurves of each variable type to derive the time series features
shown in the figure.
We also used the same subset to train the classification model for selecting MACHO QSO candidates.
For details about the training set, see Section \ref{sec:MACHO_QSO_SVM_Models}.

A simple and conventional method for selecting QSOs 
using these features is to define cuts in the 2D-space  
shown in  Figure \ref{fig:index_index} motivated by empirical observations of known classes. 
However, each panel exhibits a unique and complex structure of the features,
which suggests that defining simple cuts is difficult.
Moreover note that each panel in the figure is a 2D projection
of the original 11D time series feature space. 
This implies that even if there exist proper cuts 
in the hyperspace that can separate the classes,
these cuts could be obscured or invisible in any of the projections.
Therefore, using simple cuts empirically derived from the projection
could be inappropriate for the classification.
In order to alleviate the problem of introducing empirical cuts and 
thus to fully utilize the derived 11 time series features,
a classification algorithm should be capable of defining boundaries (e.g. cuts) in the hyperspace.
For this purpose, we employed SVM which produces hyperplanes 
between classes in any multi-dimension space.
SVM also can define non-linear boundaries using kernel functions while cuts are generally linear.
In the following section, we briefly explain SVM.

\section{Support Vector Machines}
\label{sec:Support_Vector_Machine}

SVM \citep{Boser1992} is a family of supervised machine learning algorithms
that can train a 2-class classification model using samples of two known
classes (i.e. training data).  A SVM classifier can be seen as a single node
neural network with an implicitly defined high dimensional feature space. It
is currently one of the best classification methods in machine
learning. Compared to neural networks SVM provide a flexible classification
model, avoid the problems of local minima, and reduce the need for parameter
tuning. Several efficient optimization methods have been developed for SVM
training in recent years. For an overview, discussion and practical details
the reader is referred to \citet{Cristianinic2000, BennettC00, libsvmguide}.

SVM have been applied extensively in many application areas, and in
particular to various astronomical applications such as the classification of
variable stars \citep{Wozniak2004AJb}, the selection of Active Galactic
Nuclei (AGN) candidates \citep{Zhang2004AA}, the determination of photometric
redshift \citep{Wadadekar2005PASP}, the classification of galaxies using
synthetic galaxy spectra \citep{Tsalmantza2007AA} and the morphological
classification of galaxies using image data \citep{Huertas2008AA}.

The classifier of a SVM defines a linear hyperplane that separates two
classes in a training set.  To select a unique hyperplane among the set of
possible hyperplanes that separate the data, SVM chooses the hyperplane which
maximizes the margin between the two classes, and is therefore often called
the {\em{maximum margin separator}}.
However, in many cases, it is not possible to find any hyperplane that
can perfectly separate two classes.
In other words, a training set of two classes cannot be separated without errors.
In order to solve this problem, {\em{soft margin}} SVM which allows errors 
in a training set (i.e. mislabeled samples) was proposed \citep{Cortes1995}.
The soft margin SVM uses a modified optimization criterion where a constant,
$C > 0$, controls a  
tradeoff between maximizing the margin and minimizing the errors of a classification model.
The parameter $C$ needs to be selected appropriately in every application 
to balance the margin with the errors.
Small $C$ allows a large margin between two classes and thus tends to 
ignore mislabeled samples.
On the other hand, large $C$ allows a small margin and tries to
separate even mislabeled samples.
Another approach to address non-separability is to map the examples into a
(typically high dimensional) feature space where the data might be better
separated. Such mappings are captured implicitly by SVM as well as several
other learning methods.
To achieve this, SVM employ 
non-linear kernel functions that capture inner products in the implicit
feature space. Intuitively the kernel can also be seen to be a similarity
function acting in the
expanded space.
When this is done the hypothesis of SVM has the form:

\begin{equation}
\label{eq:svmhyp}
Class(z) = \mbox{sign}(\sum_i \alpha_i y_i  K(z,x_i))
\end{equation}

{\noindent}where $z$ is the example we are predicting the label for, $x_i$ are the
training data (i.e. the vectors of time series features), 
 $y_i$ are the labels for the $i_{th}$ training data,
and $i$ are indices for training examples. 
The $\alpha_i$ are the parameters learned
by the training procedure.
The construction of SVM shows that
this form captures a linear separator in the feature space for which $K(z,x_i)$ is an
inner product, and the training procedure chooses the $\alpha_i$ that
maximize the criterion of soft margin. Despite the mapping to a potentially
high dimensional space, the maximum margin criterion leads to automatic
capacity control and thus avoids overfitting.

Many forms of kernels exist in the literature, and the the most commonly
used are the polynomial and the RBF (radial basis function) kernels.
In this work, we followed standard practice \citep{libsvmguide} and
used the RBF defined as:

\begin{eqnarray}
K(x_{i}, x_{j}) = \rm{exp}(-\gamma||x_{i}-x_{j}||^{2}),\,\,\gamma>0,
\end{eqnarray}

{\noindent}where $x_i$, $x_j$ are two examples, and the kernel parameter
$\gamma$ determines the width of the kernel function. 
The implicit feature space in this case is known to be of infinite dimension.
As in the case of the
parameter $C$, the value of $\gamma$ 
needs to be selected appropriately for the application. One can readily
observe that this kernel measures similarity between examples and that $\gamma$
controls how fast the similarity decays with respect to the distance between
the examples. Seen in this light, the classifier (Equation \ref{eq:svmhyp}) can
also be seen to be a weighted form of nearest neighbor
classification where the $\alpha_i$ weight the importance of training
examples. 

It is well known that the choice of $\gamma$ and $C$ can affect the results
dramatically. 
In order to determine the best values for our application
we used grid search with the 10-fold cross-validation and technique
 \citep{libsvmguide}.

\begin{itemize}

\item{Cross-validation}

We divide each class into 10 subsets (i.e. 10-fold cross-validation)
and select nine subsets to train a classification model.
We then apply the trained model to the remaining  subset  and count the
number of true positives (i.e. number of QSOs that the model identifies as QSOs),
the number of false positives (i.e. number of non-QSOs that the model identifies as QSOs)
and the number of  false negatives (i.e. number of QSOs that the model identifies as non-QSOs).
We repeat this process ten times with all different combinations.
Finally we sum  the true positives, false positives and false negatives
from each iteration, and calculate the recall and precision defined as:

\begin{eqnarray}
\begin{array}{l}
\displaystyle
{\rm{recall}} = \frac{N_{TP}}{N_{TP} + N_{FN}}, \,\,{\rm{precision}} = \frac{N_{TP}}{N_{TP} + N_{FP}},
\end{array}
\end{eqnarray}

{\noindent}where $N_{TP}$ is the sum of the true positives,
$N_{FP}$ is the sum of the false positives and
$N_{FN}$ is the sum of the false negatives\footnote{False positive rate is $1 - {\rm{precision}} = {N_{FP}}/({N_{TP} + N_{FP}})$}.

\item{Grid search}

To select the best $C$ and $\gamma$,
we search in  a    log-scale evenly  spaced 10x10 grid   with values
from $10^{-1}$ to $10^{4}$.
We then perform  a 10-fold cross-validation 
and select $C$ and $\gamma$ that gave the best recall and the best precision.
We then define a finer 10x10 grid and
repeat the 10-fold cross-validation test with the new set of parameters.
We repeat this procedure until recall and precision are not improving any more.

\end{itemize}

Standard SVM does not provide probability output.
Thus we employed Platt's probability estimation \citep{Platt1999} to derive class probabilities.
The Platt posterior probability is calculated using a sigmoid function as:

\begin{eqnarray}
Pr(y=1|x) = \frac{1}{1+e^{Af + B}} \,\,,
\end{eqnarray}

{\noindent}where $f$ is a decision function such that ${\rm{sgn}}(f(x))$ decides the class of sample $x$.
$y$ is the label for sample $x$ (i.e. a value for the class) and takes the
values of +1 or -1. As Platt notes, 
this amounts to assuming that $f$ corresponds to the log-odds of the positive
label; this assumption is not fully justified but has been shown to work well
in many applications. 
The parameters $A$ and $B$ are calculated by minimizing the negative
log-likelihood of a training data:

\begin{eqnarray}
\begin{array}{l}
\displaystyle
{\rm{min}} \{-\sum_{i=1}^{l}(t_{i} \, {\rm{log}}(p_{i}) + (1 - t_{i})\, {\rm{log}}(1-p_{i}))\},
\\ \\
\displaystyle
t_{i} = \frac{y_{i} + 1}{2},\,\,\,
\displaystyle
p_{i} = \frac{1}{1 + e^{Af_{i} + B}},
\end{array}
\end{eqnarray}

{\noindent}where $i$ are indices of training data, $l$ is the total number of the training data
and $y_{i}$ is a label for $i_{th}$ example.
The derived Platt class probabilities can be used to check the confidences of the predicted classes.

Many authors have studied 
feature selection methods to remove irrelevant features
(e.g. see \citealt{Blum1997, Bradley1998, Weston2001NIPS, LiICCIMA2003, Chenspringer2006} and references therein).
Such feature selections could be useful 
when there are too many features (e.g. more than a few hundred) 
including both relevant and irrelevant features.
However, \citet{Nilsson2006springer} found that most known feature selection methods occasionally 
discard even relevant features.
This work also noticed that SVM is robust against uninformative
features as long as there are a sufficient number of informative features.
Another reason for feature selection is to reduce
CPU time for extracting features and for training models 
when there exist a great number of features.
Note that we employed only 11 time series features (see the previous section and Appendix), 
and all of them are informative for separating some of the classes as shown in Figure \ref{fig:index_index}.
Thus it is not necessary  to implement feature selection methods in this work.

\section{MACHO QSO Candidate Selections Using SVM Classification Models}

\subsection{Training Classification Models}
\label{sec:MACHO_QSO_SVM_Models}

Using the 11 time series features and SVM,
we trained a classification model for selecting MACHO QSO candidates.
To train the model, we first selected a training set which consists  of 
58 MACHO QSOs\footnote{We removed one MACHO QSO from the dataset because it has only 50 data points
while the rest of the MACHO QSOs have at least several hundred data points.}, 
1,629 variable sources of known types
(128 Be stars, 582 microlensing events, 193 eclipsing binaries,
288 RR Lyraes, 73 Cepheids and 365 LPVs) 
and 4,288 non-variable sources.
We selected these variables from the list of known MACHO variable sources.
Table \ref{tab:known_MACHO_var} shows the number of the known MACHO variables
we collected from \href{http://simbad.u-strasbg.fr/simbad/}{SIMBAD}'s 
MACHO variable catalog\footnote{\href{http://vizier.u-strasbg.fr/viz-bin/VizieR?-source=II/247}{http://vizier.u-strasbg.fr/viz-bin/VizieR?-source=II/247}}
\citep{Alcock2001} and also from several literature sources 
\citep{Alcock1997ApJ, Alcock1997ApJL, Alcock1997ApJa, 
Wood2000PASA, Keller2002AJ, Thomas2005ApJ}.\footnote{\label{footnote:gabe}We 
added more than several thousands of new variable candidates selected in the MACHO LMC database to the table.
These were identified by an another group at the Time Series Center, Initiative in Innovative Computing at Harvard (\href{http://timemachine.iic.harvard.edu}{http://timemachine.iic.harvard.edu}).
The statistical characteristics of the candidates will be separately published soon.
For details about the selection algorithm, see \citet{Wachman2009}.}
To select non-variable stars, we randomly chose a subset of MACHO lightcurves from a few MACHO LMC fields
and removed all the known MACHO variables from the subset.

We then derived the 11 time series features for individual MACHO lightcurves in the training set.
Before deriving the features, we removed all data points in each lightcurve
with photometric errors greater than three times 
the average photometric errors.\footnote{SVM cannot consider errors of features while training a model.}
The photometric errors are given by the MACHO photometric pipeline \citep{Alcock1999PASP}.

\begin{table}
\begin{center}
\caption{Number of known MACHO variables\label{tab:known_MACHO_var}}
\begin{tabular}{ccc}
\tableline\tableline
Variable types & \# & References \\
\tableline
RR Lyraes  & 9,722 & \citet{Alcock2001} \\
Cepheids & 1,868 & \citet{Alcock2001} \\
Eclipsing binaries & 6,835 & \citet{Alcock2001} \\
LPVs & 3,049 & \citet{Wood2000PASA} \\

Blue variables & 1,262 & \citet{Keller2002AJ} \\
Microlensings & 626 & \citet{Alcock1997ApJ, Alcock1997ApJL, Alcock1997ApJa} \\
& & \citet{Thomas2005ApJ} \\
Be stars & 136 &  private communication \\
& & with Geha, M. \\
RR Lyraes & 8,292 & from a separate work done \\ 
Cepheids & 1,452 & by our group. \\
& & see the text for details.${}^{\ref{footnote:gabe}}$\\
\tableline
Total & 33,242 & \\
\tableline
\end{tabular}
\end{center}
\end{table}

We then employed a 2-class classification SVM\footnote{We 
used the \href{http://www.csie.ntu.edu.tw/~cjlin/libsvm}{\fontfamily{pcr}\selectfont LIBSVM package} \citep{Chang2001}.} 
using the RBF.
We empirically found that 2-class SVM with the RBF achieves better recall 
and precision than 2- or multiple-class SVM with 
other kernels including linear kernel. 
We applied a 10-fold cross-validation and grid search 
to all the combinations of 2- or multiple-class SVM and different
kernels. We found that 2-class SVM with the RBF showed the best recall and precision.
To use a 2-class SVM, we defined the MACHO QSOs as the members of one class and 
all others  as members of the other class.
In order to derive the best $C$ and $\gamma$, we performed
a 10-fold cross-validation and grid search using the training set as
described in the previous section.
We performed the test on each MACHO band; one for the B band and one for the R band.
Table \ref{tab:recall_precision} shows the derived best recall and precision of each band.
As can be seen from the table, the B (R) model shows 82.8 (72.4)\% recall and 75\% precision,
which means the B (R) model misses 17.2 (27.6)\% of the MACHO QSOs and has 25 (25)\% false positive rate.
For the B model, the false positives consist of 12 Be stars, three microlensing events and one LPV;
for the R model, 11 Be stars and three microlensing events.
Although the majority of the false positives were Be stars as expected,
the models excluded more than 90\% of the 128 Be stars in the training set. 
It is worth mentioning that recall and precision could vary depending on 
which set of variables and non-variables we choose to use as a training set.
For instance, if we exclude the 128 Be stars from the training set, 
we can increase recall to 95\% with a 7\% false positive rate.
We can further increase recall and precision 
if we also remove microlensing events and LPVs from the training set.
However, note also that the higher recall and precision 
does not guarantee a better model because the model 
would not be able to distinguish QSOs from the false positives 
such as Be stars, microlensing events and LPVs when applied to the whole dataset.

Finally, we trained two models, one each for the MACHO B and R bands,
using the derived best $C$, $\gamma$ on the whole training set\footnote{This 
model is slightly different from the one used for the cross-validation because 
it was trained on the whole training set as opposed to  9/10 of the training set.}.
We used the trained models to select QSO candidates from the MACHO database 
(see Section \ref{sec:MACHO_QSO_Candidates_Selection}).
Although the rate of derived false positives mentioned in the previous paragraph is 25\%, 
it should not be expected that 
the selected MACHO QSO candidates using the models would have 25\% false positives.
This is because the training set is not complete;
also,  it is nearly impossible to take into account
every known type of variability existing in the MACHO database,
which includes not only astronomical variables but also non-astronomical
photometric defects or systematic errors.
In addition, the fraction of QSO in the whole dataset is likely
to be different than the training set.
Thus the true false positive rate for the MACHO QSO candidates
could be higher than 25\%. We will come back to this point when we discuss
crossmatching the candidate list with known catalogs in Section \ref{sec:crossmatching}.

\begin{figure}
\begin{center}
       \includegraphics[width=0.45\textwidth]{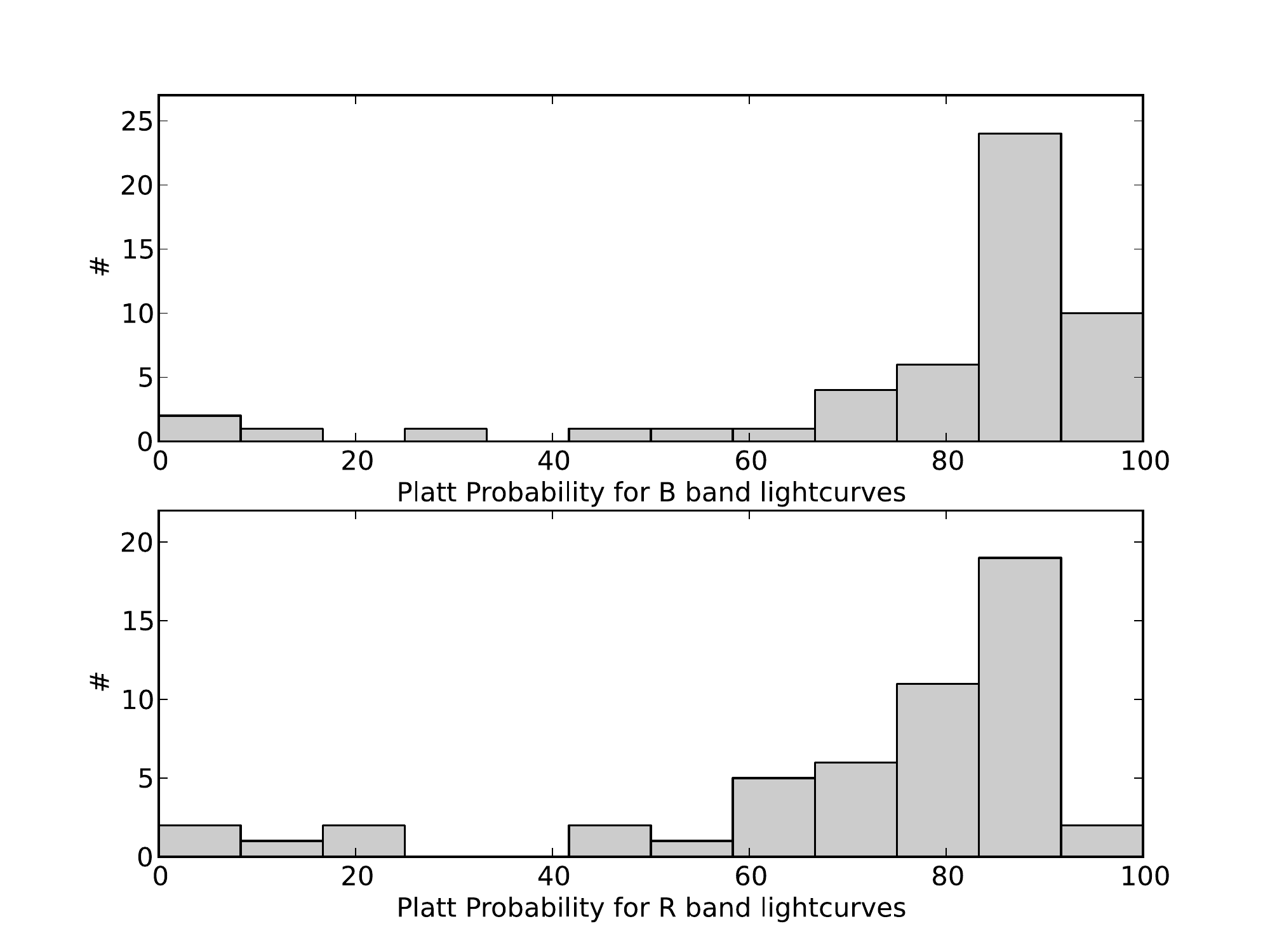} 
\end{center}
    \caption{Platt probabilities for the known MACHO QSOs. The top (bottom) panel is
    the Platt probabilities of the B (R) band lightcurves.}
    \label{fig:Platt}
\end{figure}

In addition, Figure \ref{fig:Platt} shows
the Platt probabilities of the known MACHO QSOs for 
B (the top panel) and R (the bottom panel) band lightcurves.
As the figure shows, the majority of the QSOs have higher probabilities than 80\%. 
We used the Platt probability of each MACHO lightcurve to
select MACHO QSO candidates (see Section \ref{sec:MACHO_QSO_Candidates_Selection}).

\begin{table}
\begin{center}
\caption{Recall and precision during the cross-validation\label{tab:recall_precision}}
\begin{tabular}{cccc}
\tableline\tableline
Band & Recall & Precision & False Positives\footnote{1 - Precison}\\
\tableline
B & 82.8\% & 75.0\% & 25.0\% \\
R & 72.4\% & 75.0\% & 25.0\% \\
\tableline
\end{tabular}
\end{center}
\end{table}

\subsection{MACHO QSO Candidate Selections}
\label{sec:MACHO_QSO_Candidates_Selection}

\begin{figure}
\begin{center}
       \includegraphics[width=0.45\textwidth]{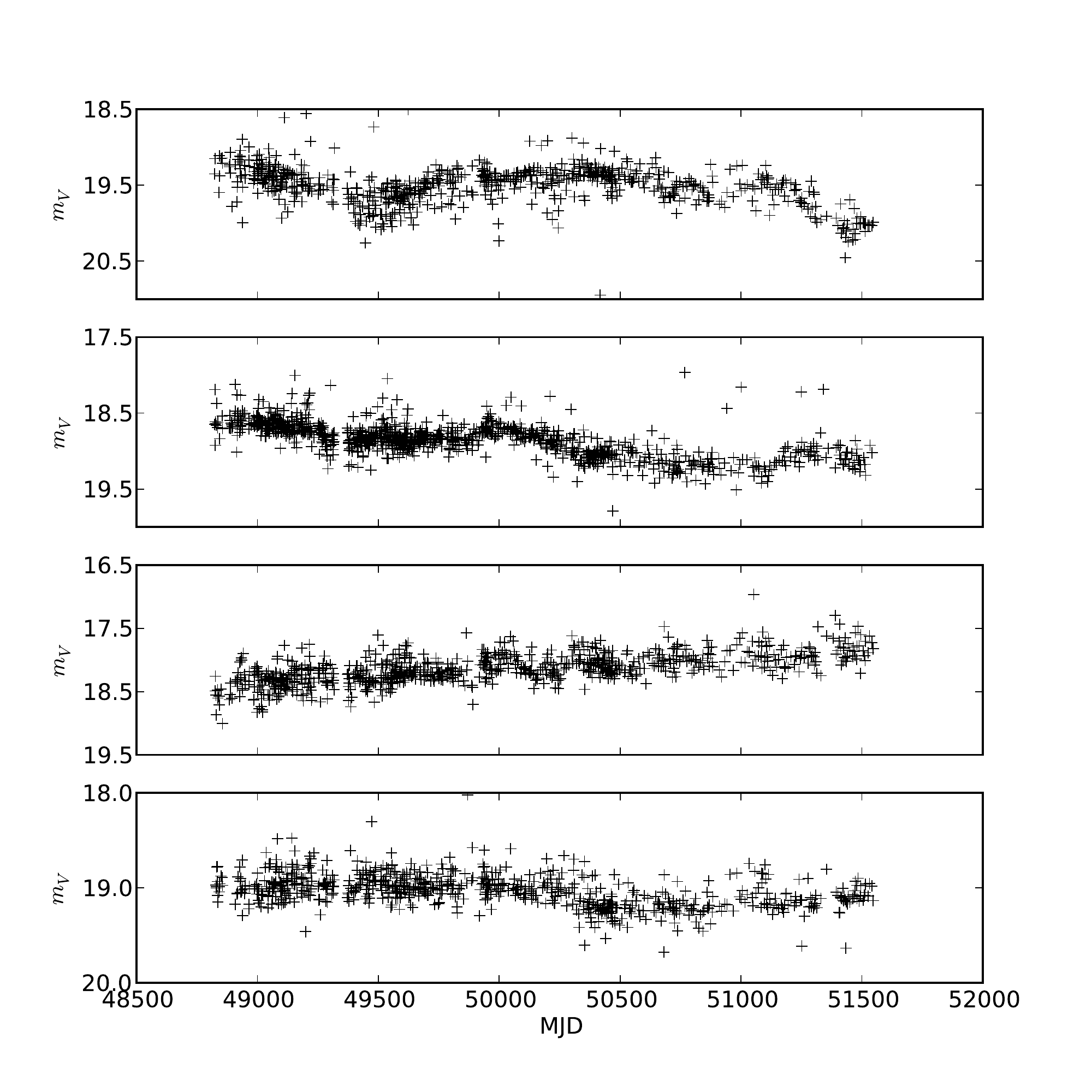}
\end{center}
    \caption{Example lightcurves of the QSO candidates. 
	The x-axis is modified Julian Date (MJD), and the y-axis is V magnitude, $m_{V}$.
    Each lightcurve manifests non-periodic and strong flux variation.}
    \label{fig:QSO_candidates}
\end{figure}

To select the MACHO QSO candidates, we first derived the 11 time series features 
for the whole 40 million MACHO LMC lightcurves.\footnote{If 
an object does not have a lightcurve of any particular band, 
we ignore that object. Nevertheless, almost all of the 20 million MACHO objects 
have both B and R band lightcurves, so the overall selection efficiency is not affected.}
We removed the data points in each lightcurve which have photometric errors greater than three times 
the average photometric errors
as we did during model training (see Section \ref{sec:MACHO_QSO_SVM_Models}).
We then applied the trained models to each lightcurve and derived the QSO Platt probability estimation.
Finally we selected only the lightcurves which had the 
probability product of B and R bands higher 
than 25\% (e.g. 50\% probabilities in both B and R bands).
Using the 25\% cut, we selected 1,620 QSO candidates from the entire  MACHO LMC database.
We show  example lightcurves of the QSO candidates in Figure \ref{fig:QSO_candidates}. 
As the figure shows, all the lightcurves have strong and non-periodic flux variation,
which is the variability characteristic of QSOs.

\begin{figure}
\begin{center}
        \includegraphics[width=0.45\textwidth]{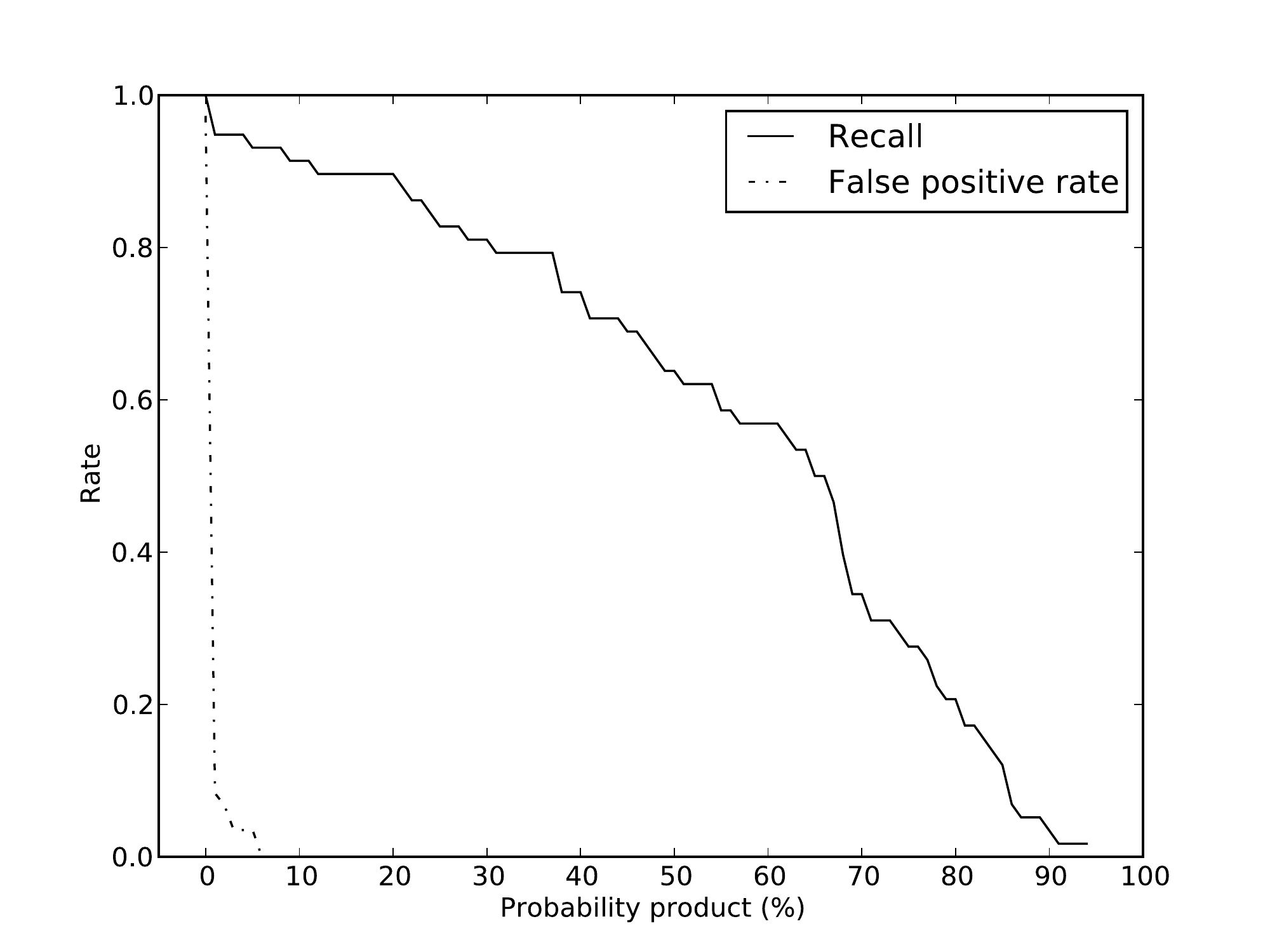}
\end{center}
    \caption{Recall and false positive rate of the models based on the training set.
    Using 25\% cut, we ca identify more than 80\% of the known MACHO QSOs
    while removing  all other variables and non-variables.}
    \label{fig:TPR_FPR}
\end{figure}

Figure \ref{fig:TPR_FPR} shows recall and false positive rates
corresponding to the probability product cuts on the training set.
Using the 25\% cut, we correctly identified
82.8\% of the known MACHO QSOs (48 out of 58) with a 0\% false positive rate.
Although a probability cut lower than 25\% yields better 
recall and also a 0\% false positive rate, we choose the 25\% cut
because our training set is not complete,
as mentioned in the previous section.

\section{Crossmatching Results with Infrared and X-ray Catalogs}
\label{sec:crossmatching}

In order to estimate the true false positive rate without spectroscopic confirmation,
we crossmatched the candidates with other astronomical catalogs.
In the following subsections, we present the crossmatching results and 
the false positive rate estimated on the basis of the crossmatched counterparts.

\subsection{Crossmatching with the Spitzer SAGE LMC catalog}
\label{sec:Spitzer_SAGE}

It is known that mid-IR color selection is efficient at separating 
AGNs from other galaxies or stars 
because the spectral energy distributions of these types
are substantially different from each other \citep{Laurent2000AA, Lacy2004ApJS, Trichas2010MNRAS, Kalfountzou2010arXiv}.
Based on these characteristics, \citet{Lacy2004ApJS} and \citet{Stern2005ApJ}
introduced a mid-IR color cut to separate AGNs
using the the Spitzer SAGE 
(Surveying the Agents of a Galaxy's Evolution; \citealt{Meixner2006AJ}) catalog.
\citet{Kozlowski2009ApJ} (hereinafter KK09) employed the mid-IR color cut
and selected about 5,000 AGN candidates from the Spitzer SAGE catalog.
KK09 also confirmed that the mid-IR color cut 
successfully identified most of the known QSOs in the SAGE footprints.

To check whether our candidates are inside the mid-IR selection cut that KK09 used,
we crossmatched them with the Spitzer SAGE LMC catalog
containing 6 million mid-IR objects and found 1,239 counterparts.
We first searched the nearest SAGE source from each of 
the candidates within a 1$^{\prime\prime}$ search radius.
In order to minimize false crossmatchings, we defined the source as a counterpart only if there exist
no other Spitzer sources within a 3$^{\prime\prime}$ radius from the candidate.

\begin{figure*}
\begin{center}
\begin{minipage}[c]{8cm}
        \includegraphics[width=1.0\textwidth]{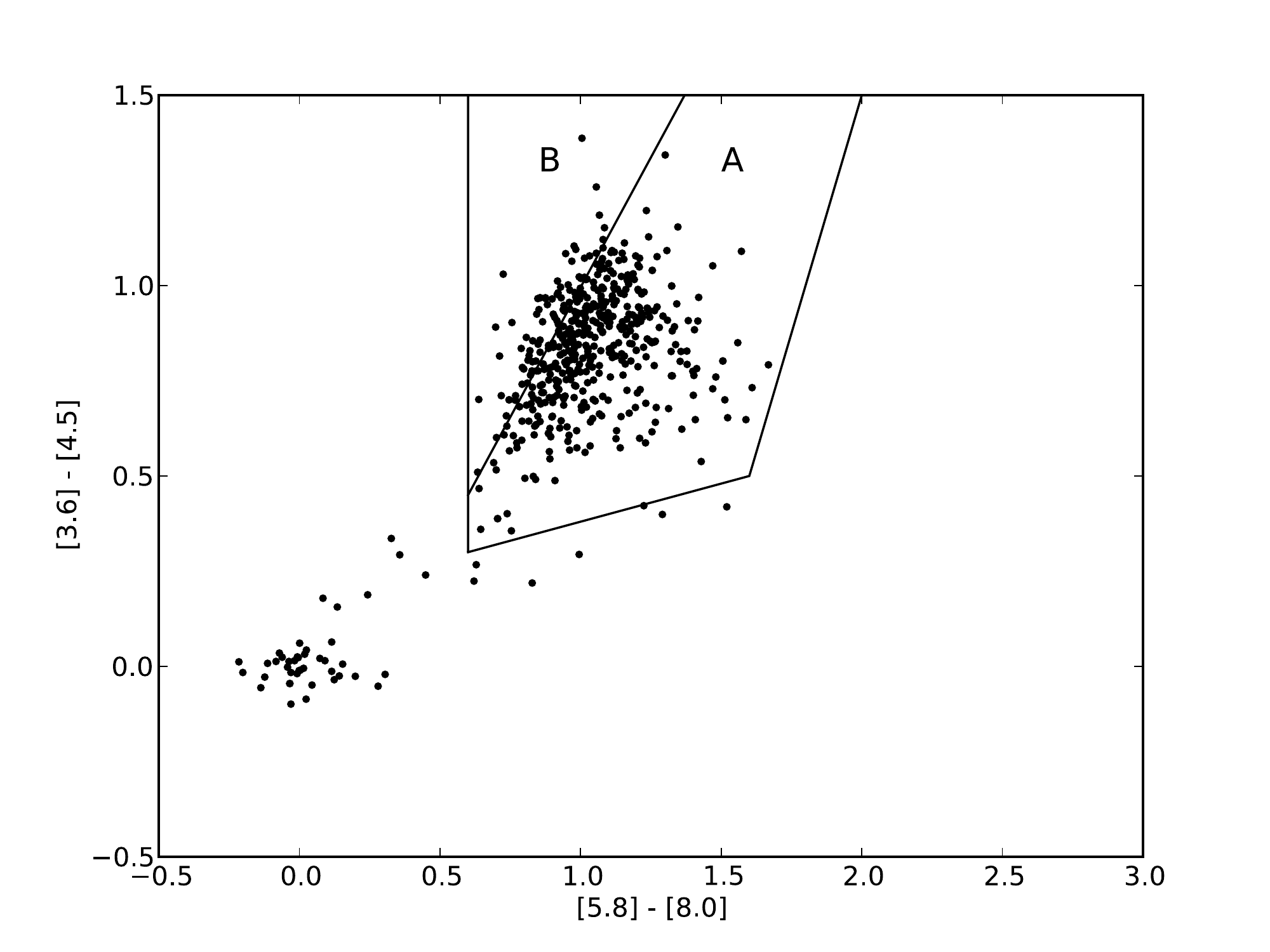}
\end{minipage}
\begin{minipage}[c]{8cm}
        \includegraphics[width=1.0\textwidth]{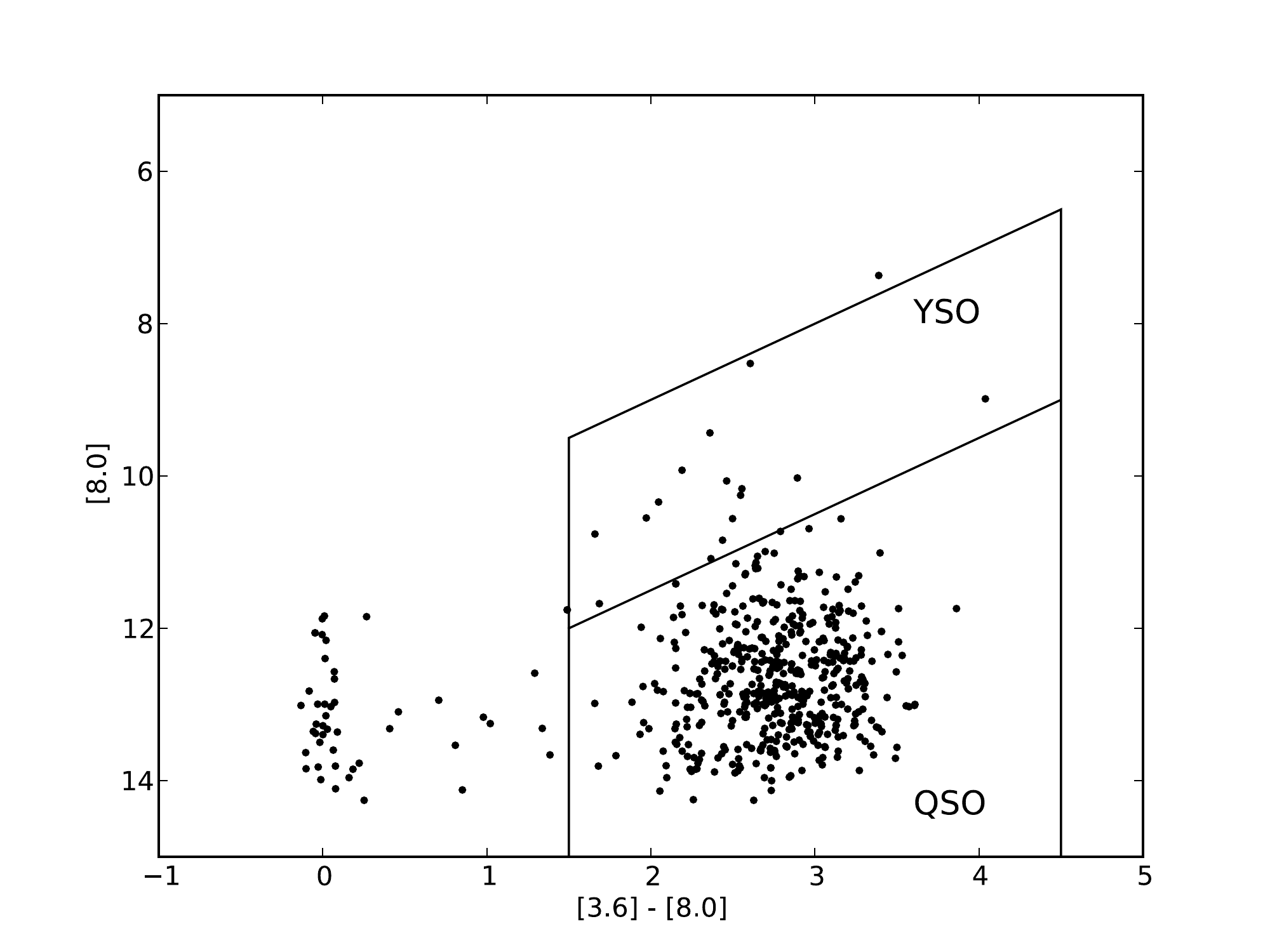}
\end{minipage}
\end{center}
    \caption
           {Mid-IR color-color and color-magnitude diagrams of 
           the Spitzer SAGE counterparts crossmatched with the QSO candidates.
           Each axis of the figure is either Spitzer magnitude or color.
           All sources inside the region A, B, QSO and YSO 
           are potential QSOs \citep{Kozlowski2009ApJ}.
           The majority of the candidates are inside the region A and QSO, 
           which is the most promising QSO regions.}
    \label{fig:sage_can_CCD_CMD}
\end{figure*}

\begin{figure}
\begin{center}
    \includegraphics[width=0.45\textwidth]{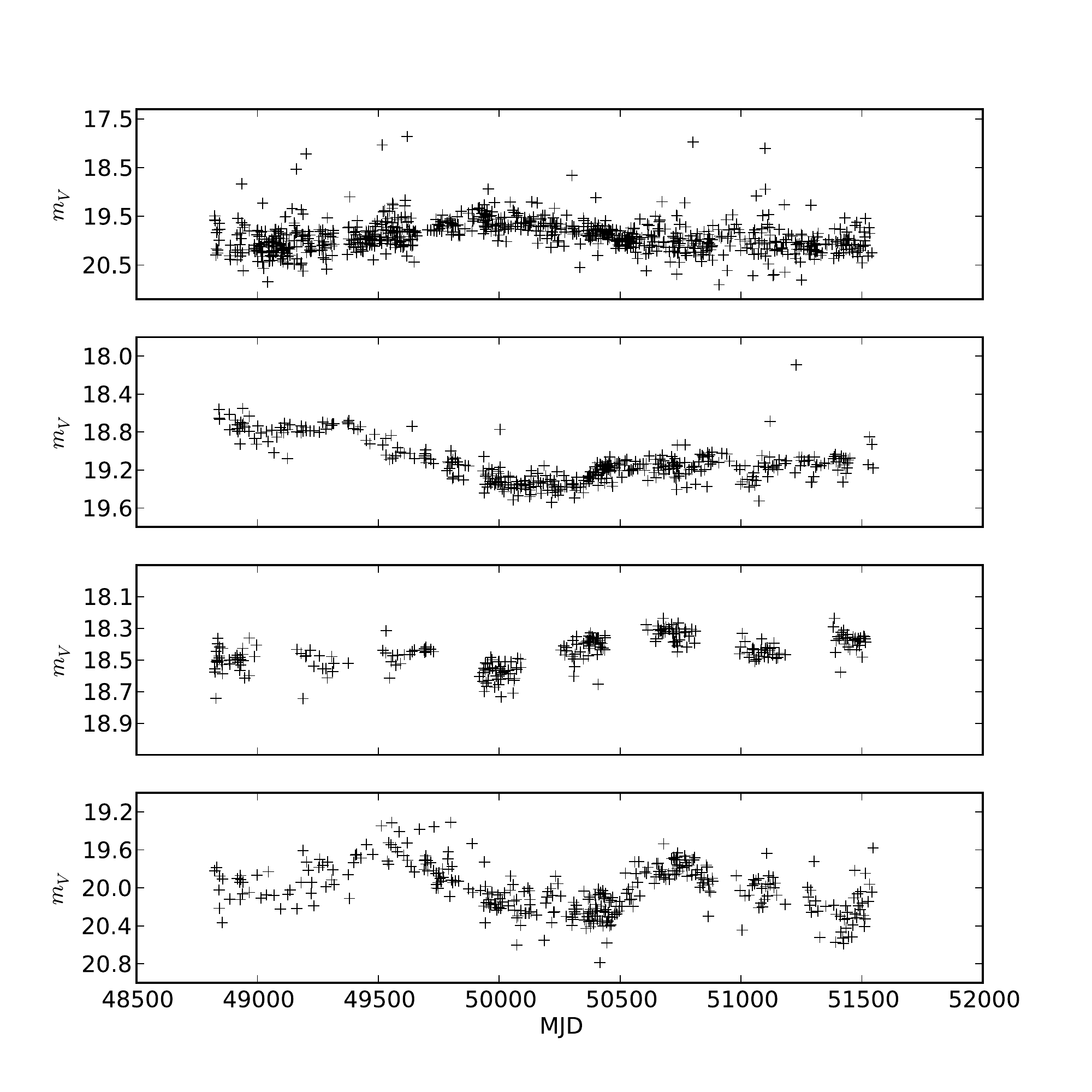}
    \caption
           {Examples lightcurves of the QSO candidates outside the AGN region.
           The x-axis is MJD, and the y-axis is V magnitude. 
           All of them show strong and non-periodic flux variation. These QSO candidates could 
           either broad or narrow emission line QSOs although they are outside the AGN region.}
    \label{fig:outside_AGN_region}
\end{center}
\end{figure}

Of the crossmatched counterparts, about 500 had 
been observed with at least three Spitzer IRAC (InfraRed Array Camera) bands.
Note that we need a minimum of three Spitzer IRAC magnitudes to apply the mid-IR color cut.
Figure \ref{fig:sage_can_CCD_CMD} shows the color-color and 
color-magnitude diagrams of these counterparts
(529 in the color-color diagram and 544 in the color-magnitude diagram).
The solid line in the figure shows the mid-IR color selection cut. 
KK09 suggested that the sources inside region B could
either be  AGNs or black bodies such as stars, 
while the sources inside region A are likely AGNs (left panel). 
In the color-magnitude diagram (right panel), there are two regions as well.
The region labeled as YSO is thought to be dominated by {\em young stellar objects} (YSO) while
the region labeled QSO is thought to be dominated by QSOs.
Nevertheless, all the sources inside these four regions (AGN region) are potential QSOs.
According to \citet{Stern2005ApJ}, the candidates inside 
the AGN region are most likely broad emission line QSOs (i.e. Type 1 AGNs).
Among them, the sources inside the QSO and A regions
are the most promising QSO candidates.
As the figure clearly shows, most of the crossmatched QSO candidates 
are inside the QSO (88.2\%; 480 out of 544) and the A regions (76.9\%; 407 out of 529),
which implies that most of the candidates are likely true QSOs.
The number of QSO candidates that are in both  the 
QSO and the A regions are 391 out of 529\footnote{529 is the total number of 
the Spitzer counterparts inside both color-color diagram and color-magnitude digram.} (73.9\%).
Under the assumption that all the 391 candidates are QSOs, the false positive rate is 26.1\%,
which is the upper bound of the false positive rate.
There are only about 9\% of the candidates outside 
the AGN region (9.3\% outside A and B regions, 9.0\% outside YSO and QSO regions), 
giving us the lower bound of the false positive rate.
Nevertheless, we confirmed that most
of the candidates outside the AGN region also show strong variability.
We show example lightcurves of these candidates in Figure \ref{fig:outside_AGN_region}.
As the figure shows, they have strong and non-periodic flux variation.
Note that our method used variability characteristics of lightcurves in order to select QSO candidates
which could be missed by the mid-IR color selection.
Moreover, the mid-IR color cut is not very efficient at 
selecting narrow emission line QSOs \citep{Stern2005ApJ}.
Therefore some of the candidates could be either
broad or narrow emission line QSOs
even though they are not inside the AGN region,
which would further decrease the lower bound of the false positive rate.

\begin{figure*}
\begin{center}
\begin{minipage}[c]{8cm}
        \includegraphics[width=1.0\textwidth]{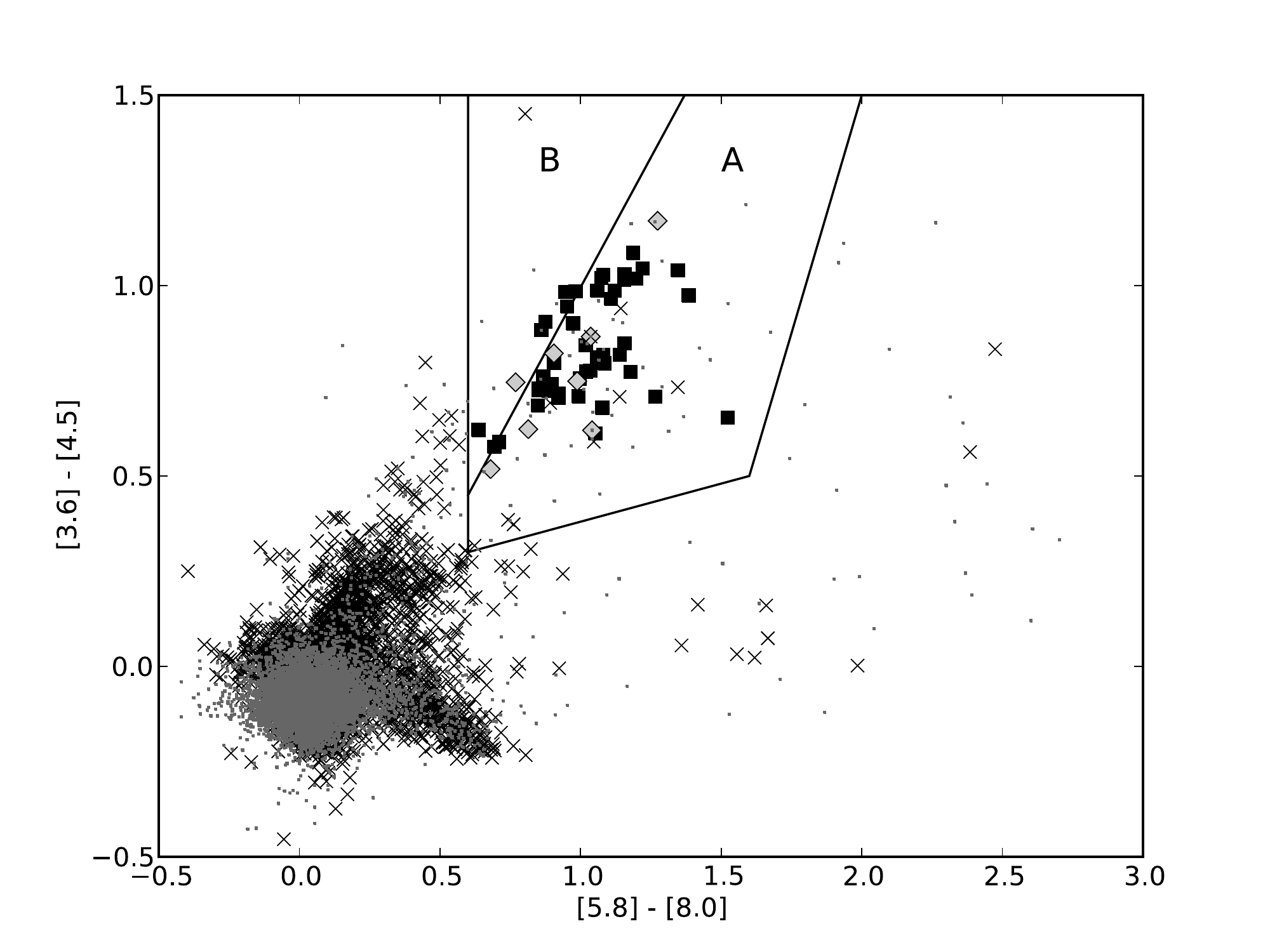}
\end{minipage}
\begin{minipage}[c]{8cm}
        \includegraphics[width=1.0\textwidth]{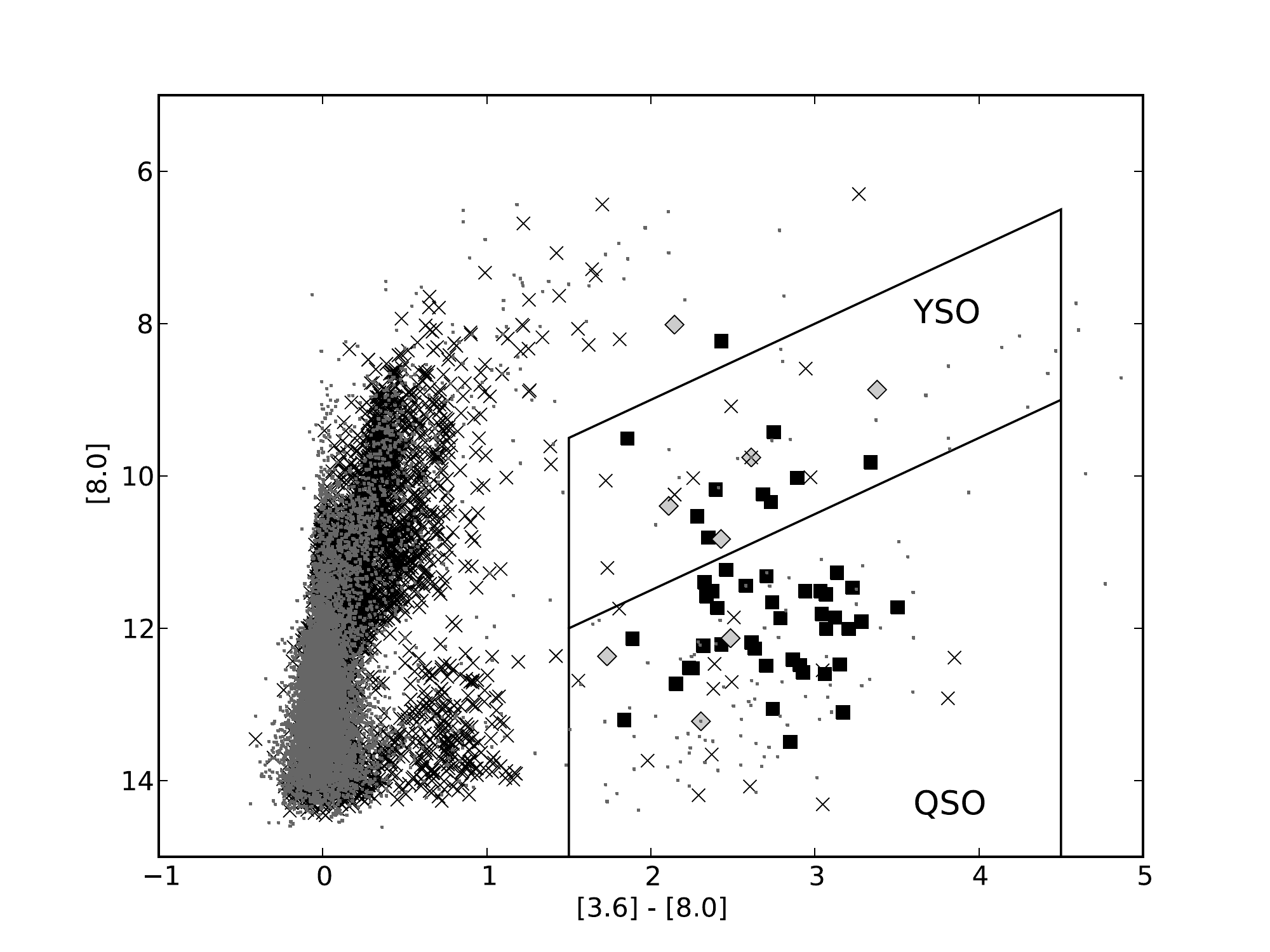}
\end{minipage}
\end{center}
    \caption
           {Mid-IR colors of the Spitzer SAGE counterparts with
           the known MACHO variable stars and the MACHO field sources.
           Each axis of the figure is either Spitzer magnitude or color.
           The black squares are MACHO QSOs, the gray diamonds are Be stars,
	the black crosses are variable stars including RR Ryraes, Cepheids,
	eclipsing binaries, LPVs and blue variable stars.
	The gray dots are MACHO field sources.
	Almost all MACHO QSOs are inside the region A, B, QSO
	and YSO, which indicates the mid-IR selection criteria is efficient at 
	confirming QSOs. However there are a lot of other variable stars including Be stars inside the regions as well.
	Thus the mid-IR selection might not be practical for selecting QSO candidates.
}
    \label{fig:sage_var_CCD_CMD}
\end{figure*}

In addition, we also crossmatched the known MACHO QSOs and the 33,242 MACHO variables shown in  
Table \ref{tab:known_MACHO_var} with the SAGE catalog 
to check how many known MACHO QSOs and known variables are inside the AGN region.
Such variables inside the AGN region could be contaminants (i.e. false positives) for any mid-IR color selection method.
We found about 50 counterparts with the known 
MACHO QSOs and about 3,900 counterparts with the variables.
We also crossmatched about 200,000 MACHO field sources from one
randomly selected MACHO field with the SAGE catalog and found $\sim$10,000 counterparts.
These field source counterparts might consist of all types of objects including non-variable stars, 
unclassified variable stars and galaxies.
Figure \ref{fig:sage_var_CCD_CMD} shows all the crossmatched counterparts.
The black squares are the MACHO QSO counterparts
(48 in the color-color diagram and 49 in the color-magnitude diagram).
The black crosses are the counterparts with the variables including RR Lyraes, Cepheids,
eclipsing binaries, LPVs and blue variable stars
(3,871 in the color-color diagram and 3,880 in the color-magnitude diagram).
We separately depict eight Be stars as gray diamonds in the figure. 
The gray dots are the MACHO field source counterparts 
(10,238 in the color-color diagram and 10,292 in the color-magnitude diagram).
As the figure shows, almost all of the MACHO QSOs are inside the AGN region as expected.
However, a few tens of the variables and 
the MACHO field sources are also inside the AGN region.
We checked these variables in the AGN region and
found that they consist of all types of the known MACHO variable stars
such as RR Lyraes, Cepheids, eclipsing binaries, blue variables and LPVs.
Moreover nearly all Be stars that have Spitzer counterparts are inside the region as well.
It is known that Be stars are characterized by their IR emission
due to dusty circumstellar environments \citep{Malfait1998AA, Leinert2004AA}.
Also note that we crossmatched only 200,000 MACHO field sources with the Spitzer catalog. 
If we scaled our selection to the total MACHO LMC database covering 20 million stars,
more than several thousand field sources would be in the AGN region,
providing significant contaminantion for QSO selection.
According to the results, it seems that the mid-IR cut is not efficient
for separating QSO candidates from various types of stars 
although it is practical for confirming QSO candidates,
especially when applied to massive databases.
In other words, the mid-IR selection cut shows
relatively low precision, although it shows high recall.
Thus it is clear that algorithms based on the
variability of lightcurves, including ours, are important
for QSO candidate selections.

\subsection{Crossmatching with X-ray Catalogs}

We crossmatched our QSO candidates with the Chandra X-ray source catalog \citep{Evans2009} and
XMM-Newton 2$^{nd}$ Incremental Source catalog \citep{Watson2009AA}.
We searched for the nearest Chandra (XMM) source
within a 5$^{\prime\prime}$ search radius from the candidate.
We only selected the source as a counterpart if there existed
no other Chandra (XMM) sources within the search radius.
Nevertheless, most of the X-ray counterparts were placed within
a 3$^{\prime\prime}$  distance from the candidates.

As a result, we found 60 X-ray counterparts.
It is known that QSOs show higher X-ray to optical flux ratios  
than typical galaxies or stars, $f_{X}/f_{r}$, owing to the accretion on 
the central black holes \citep{Reeves2000MNRAS, Hornschemeier2001ApJ}. 
To calculate  $f_{X}/f_{r}$, we first derived the $m_{V}$ and $m_{R}$ 
(i.e. standard Johnson's $V$ and Kron$-$Cousins $R$)
using the MACHO B and R magnitudes \citep{Alcock1999PASP, Kunder2008AJ}.
We then converted the $m_{V}$ and $m_{R}$ to SDSS $r$ magnitude using the formula from the
\href{http://www.sdss.org/dr4/algorithms/sdssUBVRITransform.html#Lupton2005}{SDSS website}\footnote{\href{http://www.sdss.org/dr4/algorithms/sdssUBVRITransform.html\#Lupton2005}{http://www.sdss.org/dr4/algorithms/sdssUBVRITransform.html}} \citep{Lupton2005AAS}.
Note that this formula was derived not based not on QSOs but on
photometric standard stars \citep{Stetson2000PASP}.
Thus the converted SDSS $r$ magnitudes of QSOs could 
have larger errors (i.e. standard deviation)
than the estimated errors for the standard stars, $\sigma \simeq 0.01$.
Nevertheless, we finally used the following equation from \citet{Green2004ApJS} to derive  ${\rm{log}}(f_{X}/f_{r})$:

\begin{eqnarray}
\label{eq:xray_to_optical}
\displaystyle {\rm{log}} \frac{f_{X}}{f_{r}} = {\rm{log}}\,f_{X} + 0.4\,r + 5.67,
\end{eqnarray}

{\noindent}where $f_{X}$ is the X-ray flux in units of ergs cm$^{-2}$ s$^{-1}$
in the range of 0.5-2.0 keV, 
which is extracted from the Chandra and XMM catalogs\footnote{ergs cm$^{-2}$ s$^{-1}$, which
is the unit for the Chandra sources, is identical with 10$^{-3}$ Watt m$^{-2}$, 
which is the unit for the XMM sources.}.
$f_{r}$ is the optical flux and $r$ is the converted SDSS $r$ magnitude.

\begin{figure}
\begin{center}
       \includegraphics[width=0.45\textwidth]{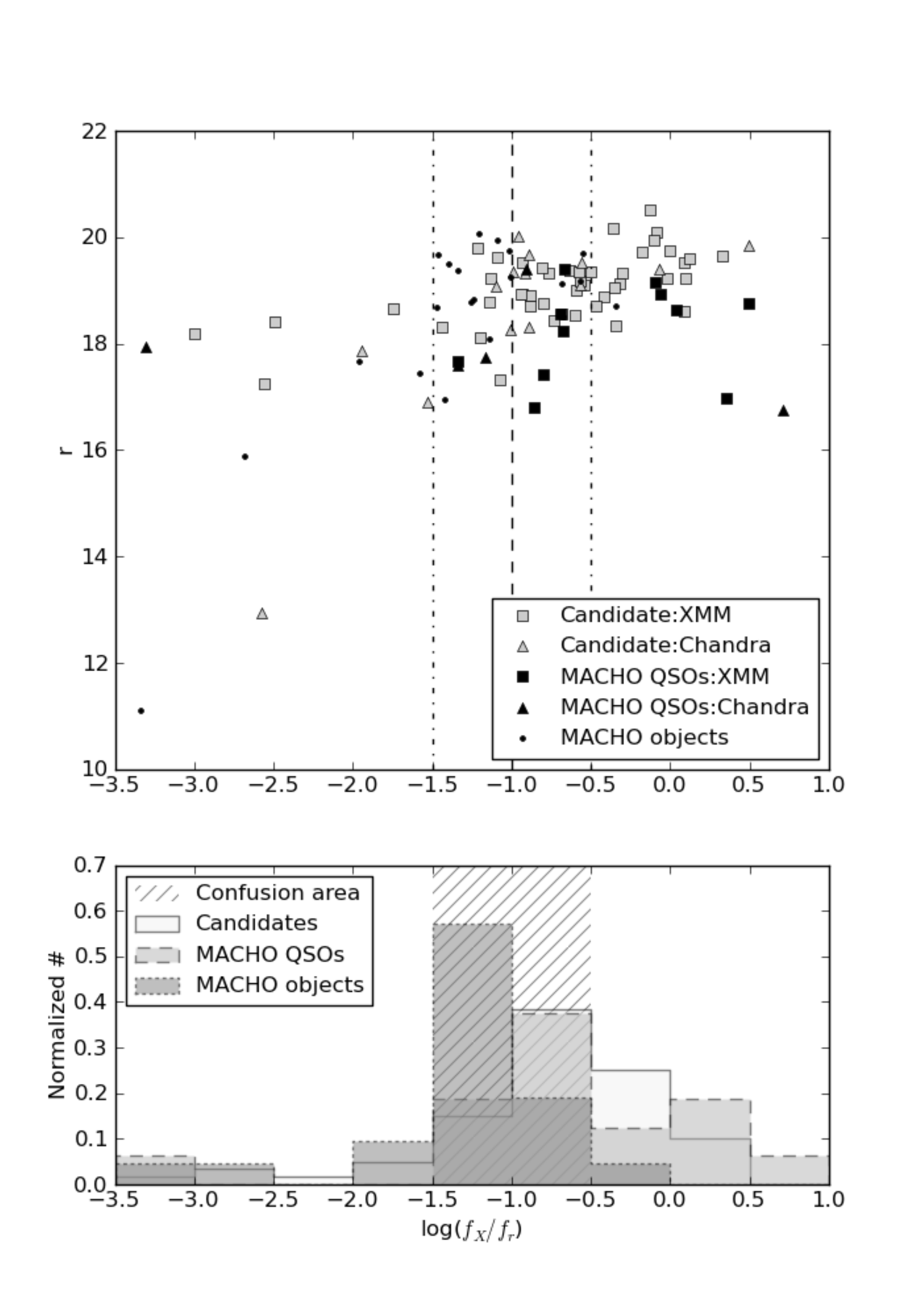} 
\end{center}
    \caption{$f_{X}/f_{r}$ of the X-ray counterparts with the MACHO QSOs, the QSO candidates and the MACHO field objects.
The top panel: the x-axis is log($f_{X}/f_{r}$), and the y-axis is the converted SDSS $r$ magnitude.
The squares are the XMM counterparts and the triangles are the Chandra counterparts.
The black marks are the MACHO QSOs and the gray marks are the candidates.
The gray dots are the MACHO field objects. 
The dashed line is the criterion between AGN and others such as galaxies and stars.
Most of the MACHO QSOs and the candidates have
higher $f_{X}/f_{r}$ than the criterion while most of the MACHO field objects
have smaller $f_{X}/f_{r}$ than the criterion, which implies most of the candidates 
are promising QSO candidates.
The bottom panel: the histogram of $f_{X}/f_{r}$.
The x-axis is log($f_{X}/f_{r}$), and the y-axis is normalized count.
Based on the histogram, we refined the region of confidence. See the text for details.}
    \label{fig:xray_optical}
\end{figure}

The top panel of Figure \ref{fig:xray_optical} shows the $f_{X}/f_{r}$
of 60 counterparts with the Chandra and XMM catalogs.
The x-axis is ${\rm{log}}(f_{X}/f_{r})$, and the y-axis is the converted $r$ magnitude. 
In the panel, we also show 16  known MACHO QSOs that have X-ray counterparts.
The black marks are the MACHO QSO counterparts, and the gray marks are
the QSO candidate counterparts. 
The squares are XMM counterparts, and the triangles are Chandra counterparts.
The dashed line corresponds to  $f_{X}/f_{r} = 0.1$,
which is the criterion separating AGNs and typical galaxies or stars \citep{Green2004ApJS}.
The two dash-dotted lines are boundaries of the confusion area 
shown as the dashed area in the bottom panel (see the following paragraph).
As the figure shows, most of the MACHO QSOs (75.0\%; 12 out of 16)
and our QSO candidates (73.3\%; 44 out of 60) show higher $f_{X}/f_{r}$ than 0.1. 
If all the candidates with higher $f_{X}/f_{r}$ than 0.1
are QSOs, the false positive rate is 27.3\%.

\begin{figure}
\begin{center}
       \includegraphics[width=0.45\textwidth]{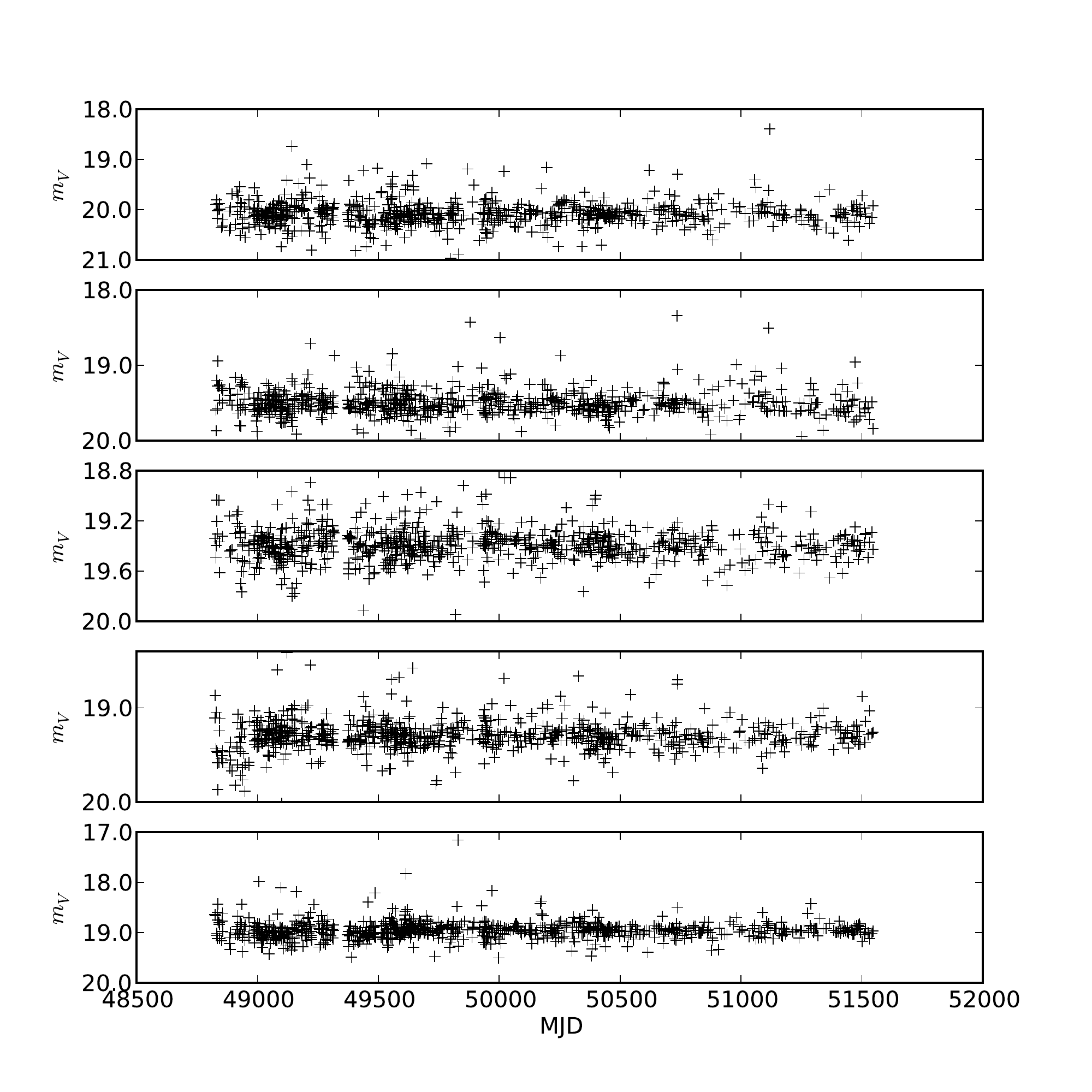}
\end{center}
    \caption{Lightcurves of 5 MACHO field objects which have 
     higher $f_{X}/f_{r}$ than the AGN criterion.
    The x-axis is MJD, and the y-axis is V magnitude, $m_{V}$.
    Although they have higher $f_{X}/f_{r}$ than the criterion, 
    they do not have strong flux variation and
    thus were not selected as QSO candidates by our selection algorithm.
    }
    \label{fig:xray_stars}
\end{figure}

In addition, to estimate how a large portion of non-AGNs could have $f_{X}/f_{r} \geq 0.1$,
we crossmatched all the objects 
from one MACHO field with the Chandra X-ray catalog.
We selected the field so that it overlapped with the Chandra footprints.
In the top panel of Figure \ref{fig:xray_optical}, 
we show the $f_{X}/f_{r}$ of the 21 crossmatched MACHO objects (black dots).
These counterparts could be either stars or AGNs, although they are most likely X-ray emitting stars
such as X-ray binaries, W-UMa binaries \citep{Chen2006AJ}, 
Algol type binaries \citep{Singh1995ApJ} and
cataclysmic variable stars (e.g. see \citealt{Wonnacott1994MNRAS})
since the number density of such stars surpasses the number density of AGNs.
Of the 21 MACHO objects, 16) have $f_{X}/f_{r}$ smaller than 0.1,
which implies that non-AGN objects generally have smaller $f_{X}/f_{r}$ than 0.1.
The remaining five objects have $f_{X}/f_{r}$ larger than 0.1 and could be AGN candidates.
We show the lightcurves of these five objects in Figure \ref{fig:xray_stars}.
As the figure shows, they do not manifest any strong flux variation
and thus were not selected as QSO candidates by our selection method.

Based on the crossmatching results mentioned in the previous paragraphs,
we further improved the region of confidence
using the histogram of ${\rm{log}}(f_{X}/f_{r})$ shown in the bottom panel of Figure \ref{fig:xray_optical}.
The x-axis is ${\rm{log}}(f_{X}/f_{r})$, and the y-axis is normalized count. 
The solid line with light gray is the histogram of the QSO candidate counterparts,
the dashed line with medium gray is the histogram of the MACHO QSO counterparts
and the dotted line with dark gray is the histogram of the 21 MACHO object counterparts.
The dashed area shows the confusion area where stars and QSOs could be mixed together.
Considering all the histograms, we modified the confidence regions as:

\begin{itemize}

\item{log$(f_{X}/f_{r}) < -1.5$} : the non-QSO area

In this region, log($f_{X}/f_{r}$) is much smaller than the AGN criterion of log$(f_{X}/f_{r})=-1$.
Thus the candidates in this region are not likely QSOs.
There are only 1 out of 16 (6.2\%) MACHO QSOs, 4 out of 21 (19\%)  MACHO objects, 
7 out of the 60 (11\%) candidates inside this region.

\item{$-1.5 \leq f_{X}/f_{r} < -0.5$} :  the confusion area that is a mixture of stars and QSOs

Most of the MACHO objects (76.2\%; 16 out of 21) are in this region.
More than a half of the MACHO QSOs (55.3\%; 9 out of 16) 
and 32 out of the 60 QSO candidates (53.3\%) are also in this region.

\item{$f_{X}/f_{r} \geq -0.5$} : the QSO area

Most of the candidates in this region would be 
QSOs because of their high $f_{X}/f_{r}$.
As the histogram shows,  only 1 out of 21 (5\%) MACHO objects is in this region
while 6 out of 16 (37.5\%) MACHO QSOs and 21 out of 60 (35\%) candidates are inside the region.
\end{itemize}

As we mentioned above, 21 out of the 60 candidates are inside the QSO area and are likely true QSOs, 
which gives the upper bound of the false positive rate, 65.0\% (39/60).
In addition, some of the 32 candidates inside the confusion area could be also QSOs 
because more than half of the known MACHO QSOs are inside the confusion area. 
Thus the lower bound of the false positive rate is 11.7\% (7/60).

\section{Ongoing and Future Works}
\label{sec:Future_Works}

We will observe the QSO candidates with spectroscopic instruments 
to check whether they are QSOs.
Based on the projection of the models and the crossmatching results,
we expect at least several hundred  candidates to turn out to be  QSOs.

Using the confirmed QSOs and the false positives, we will improve our model.
The current model is constructed based on the 
relatively small number of known QSOs (i.e. 58 known MACHO QSOs),
which may be too small a sample to represent the true variability
characteristics of all QSOs in the MACHO database.
Thus using a large number of QSOs (i.e. more than a few hundreds)  would help improve the models.

In addition, our model is effective  at selecting
not only QSOs but also other types of variable sources.
Preliminary tests showed that recall and precision for periodic variables such as RR Lyraes, 
Cepheids and eclipsing binaries, were almost 100\%;
for LPVs, microlensing events and Be stars, recall and precision were 80\%.

\section{Summary}
\label{sec:Summary}

In this paper, we presented a new QSO selection algorithm
based on 11 time series features 
and a supervised classification.
We first introduced 11 time series features to quantify 
variability characteristics of lightcurves.
We then used Support Vector Machine (SVM) to train a classification model
which separates QSOs from other types of variable stars and non-variable stars.
Using the training set of the MACHO variables
(128 Be stars, 582 microlensing events, 193 eclipsing binaries, 288 RR Lyraes, 73 Cepheids and 365 LPVs), 
4,288 non-variables and the 58 known MACHO QSOs, 
we trained the models for each MACHO B and R band.
The trained model correctly identified about 80\% of the MACHO
QSOs with 25\% false positive rates on a cross-validation test.
The majority of false positives during the training were Be stars known to show variability similar to with QSOs.

We applied the model to the whole MACHO LMC database 
consisting of 40 million lightcurves (i.e. 20 million from each MACHO band)
in order to select QSO candidates. As a result, we found 1,620 candidates from the MACHO LMC database.
During the selection, none of the known MACHO variables
were mis-selected as QSO candidates.
To estimate the true false positive rate of the QSO candidates,
we crossmatched the candidates with 
astronomical catalogs, including the Spitzer SAGE LMC catalog and some X-ray catalogs.
The crossmatching results confirmed that most of our candidates are promising QSO candidates. 
For instance, the majority of candidates with Spitzer counterparts are inside the AGN region 
that is defined by a mid-IR color cut and is known to be effective in  
confirming QSO candidates. 
The crossmatching with X-ray catalogs shows that  
most of the X-ray counterparts have  $f_{X}/f_{r} \geq 0.1$ and therefore 
are likely QSOs.

In addition, during the crossmatching with the SAGE LMC catalog,
we found that using only the mid-IR color cut is not a very efficient method
in selecting QSO candidates, although it is an effective method  in confirming  QSOs.
This suggests that selection methods using variability 
characteristics of lightcurves, including ours, are important
to further remove false positives, both variables and non-variables.

\section*{Acknowledgements}

We thank M.-S. Shin,
J. D. Hartman, P. J. Green and
R. Reid
for comments and helpful advice.
The analysis in this paper has been done using the \href{http://hptc.fas.harvard.edu/}{Odyssey cluster} supported by the FAS Research 
Computing Group at \href{http://harvard.edu/}{Harvard}.
This work has been supported by NSF grant IIS-0803409.
This research has made use of the \href{http://simbad.u-strasbg.fr/simbad/}{SIMBAD} database, operated at CDS, Strasbourg, France.


\vspace{0.1cm}
\bibliography{Kim2011_QSOs}{}

\begin{thebibliography}{107}
\expandafter\ifx\csname natexlab\endcsname\relax\def\natexlab#1{#1}\fi

\bibitem[{{Alcock} {et~al.}(2001){Alcock}, {Allsman}, {Alves}, \& {et
  al.}}]{Alcock2001}
{Alcock}, C., {Allsman}, R., {Alves}, D.~R., \& {et al.} 2001, Variable Stars
  in the Large Magellanic Clouds, VizieR Online Data Catalog
  (http://vizier.u-strasbg.fr/viz-bin/VizieR?-source=II/247)

\bibitem[{{Alcock} {et~al.}(1997{\natexlab{a}}){Alcock}, {Allsman}, {Alves}, \&
  {et al.}}]{Alcock1997ApJL}
{Alcock}, C., {Allsman}, R.~A., {Alves}, D.~R., \& {et al.} 1997{\natexlab{a}},
  ApJ, 491, L11+

\bibitem[{{Alcock} {et~al.}(1997{\natexlab{b}}){Alcock}, {Allsman}, {Alves}, \&
  {et al.}}]{Alcock1997ApJa}
---. 1997{\natexlab{b}}, ApJ, 479, 119

\bibitem[{{Alcock} {et~al.}(1997{\natexlab{c}}){Alcock}, {Allsman}, {Alves}, \&
  {et al.}}]{Alcock1997ApJ}
---. 1997{\natexlab{c}}, ApJ, 486, 697

\bibitem[{{Alcock} {et~al.}(1999){Alcock}, {Allsman}, {Alves}, \& {et
  al.}}]{Alcock1999PASP}
---. 1999, PASP, 111, 1539

\bibitem[{{Alcock} {et~al.}(2000){Alcock}, {Allsman}, {Alves}, \& {et
  al.}}]{Alcock2000ApJ}
---. 2000, ApJ, 542, 281

\bibitem[{{Alcock} {et~al.}(1996){Alcock}, {Allsman}, {Axelrod}, \& {et
  al.}}]{Alcock1996ApJ}
{Alcock}, C., {Allsman}, R.~A., {Axelrod}, T.~S., \& {et al.} 1996, ApJ, 461,
  84

\bibitem[{{Aretxaga} {et~al.}(1997){Aretxaga}, {Cid Fernandes}, \&
  {Terlevich}}]{Aretxaga1997MNRAS}
{Aretxaga}, I., {Cid Fernandes}, R., \& {Terlevich}, R.~J. 1997, MNRAS, 286,
  271

\bibitem[{{Bauer} {et~al.}(2009){Bauer}, {Baltay}, {Coppi}, \& {et
  al.}}]{Bauer2009ApJ}
{Bauer}, A., {Baltay}, C., {Coppi}, P., \& {et al.} 2009, ApJ, 696, 1241

\bibitem[{{Becker} {et~al.}(2001){Becker}, {Fan}, {White}, \& {et
  al.}}]{Becker2001AJ}
{Becker}, R.~H., {Fan}, X., {White}, R.~L., \& {et al.} 2001, AJ, 122, 2850

\bibitem[{Bennett \& Campbell(2000)}]{BennettC00}
Bennett, K.~P., \& Campbell, C. 2000, SIGKDD Explorations, 2, 1

\bibitem[{{Blanco} \& {Heathcote}(1986)}]{Blanco1986PASP}
{Blanco}, V.~M., \& {Heathcote}, S. 1986, PASP, 98, 635

\bibitem[{{Blum} \& {Langley}(1997)}]{Blum1997}
{Blum}, A.~L., \& {Langley}, P. 1997, Artificial Intelligence, 97, 245 ,
  relevance

\bibitem[{{Boser} {et~al.}(1992){Boser}, {Guyon}, \& {Vapnik}}]{Boser1992}
{Boser}, B.~E., {Guyon}, I.~M., \& {Vapnik}, V.~N. 1992, in Proceedings of the
  fifth annual workshop on Computational learning theory, COLT '92 (New York,
  NY, USA: ACM), 144--152

\bibitem[{{Bradley} \& {Mangasarian}(1998)}]{Bradley1998}
{Bradley}, P.~S., \& {Mangasarian}, O.~L. 1998, in ICML '98: Proceedings of the
  Fifteenth International Conference on Machine Learning (San Francisco, CA,
  USA: Morgan Kaufmann Publishers Inc.), 82--90

\bibitem[{{Butler} \& {Bloom}(2010)}]{Butler2010}
{Butler}, N.~R., \& {Bloom}, J.~S. 2010, ArXiv e-prints

\bibitem[{{Chang} \& {Lin}(2001)}]{Chang2001}
{Chang}, C.~C., \& {Lin}, C.~J. 2001, LIBSVM : a library for support vector
  machines, http://www.csie.ntu.edu.tw/$\sim$cjlin/libsvm

\bibitem[{{Chen} {et~al.}(2006){Chen}, {Sanchawala}, \& {Chiu}}]{Chen2006AJ}
{Chen}, W.~P., {Sanchawala}, K., \& {Chiu}, M.~C. 2006, AJ, 131, 990

\bibitem[{{Chen} \& {Lin}(2006)}]{Chenspringer2006}
{Chen}, Y.-W., \& {Lin}, C.-J. 2006, in Studies in Fuzziness and Soft
  Computing, Vol. 207, Feature Extraction, ed. I.~Guyon, M.~Nikravesh, S.~Gunn,
  \& L.~Zadeh (Springer Berlin, Heidelberg), 315--324

\bibitem[{{Cortes} \& {Vapnik}(1995)}]{Cortes1995}
{Cortes}, C., \& {Vapnik}, V. 1995, Machine Learning, 20, 273

\bibitem[{{Cristiani} {et~al.}(1996){Cristiani}, {Trentini}, {La Franca}, \&
  {et al.}}]{Cristiani1996AA}
{Cristiani}, S., {Trentini}, S., {La Franca}, F., \& {et al.} 1996, A\&A, 306,
  395

\bibitem[{{Cristianini} \& {Shawe-Taylor}(2000)}]{Cristianinic2000}
{Cristianini}, N., \& {Shawe-Taylor}, J. 2000, An Introduction to Support
  Vector Machines (Cambridge: Cambridge Univ. Press)

\bibitem[{{de Vries} {et~al.}(2005){de Vries}, {Becker}, {White}, \& {et
  al.}}]{DeVries2005AJ}
{de Vries}, W.~H., {Becker}, R.~H., {White}, R.~L., \& {et al.} 2005, AJ, 129,
  615

\bibitem[{{Dobrzycki} {et~al.}(2005){Dobrzycki}, {Eyer}, {Stanek}, \& {et
  al.}}]{Dobrzycki2005AA}
{Dobrzycki}, A., {Eyer}, L., {Stanek}, K.~Z., \& {et al.} 2005, A\&A, 442, 495

\bibitem[{{Dobrzycki} {et~al.}(2002){Dobrzycki}, {Groot}, {Macri}, \& {et
  al.}}]{Dobrzycki2002ApJ}
{Dobrzycki}, A., {Groot}, P.~J., {Macri}, L.~M., \& {et al.} 2002, ApJ, 569,
  L15

\bibitem[{{Ellaway}(1978)}]{Ellaway1978}
{Ellaway}, P. 1978, Electroencephalography and Clinical Neurophysiology, 45,
  302

\bibitem[{{Evans} {et~al.}(2009){Evans}, {Primini}, {Glotfelty}, \& {et
  al.}}]{Evans2009}
{Evans}, I., {Primini}, F.~A., {Glotfelty}, K.~J., \& {et al.} 2009, in
  Chandra's First Decade of Discovery, Proceedings of the conference held 22-25
  September, 2009 in Boston, MA. Edited by Scott Wolk, Antonella Fruscione, and
  Douglas Swartz, abstract \#94, ed. {S.~Wolk, A.~Fruscione, \& D.~Swartz}

\bibitem[{{Eyer}(2002)}]{Eyer2002AcA}
{Eyer}, L. 2002, Acta Astronomica, 52, 241

\bibitem[{{Fan} {et~al.}(2006){Fan}, {Strauss}, {Becker}, \& {et
  al.}}]{Fan2006AJ}
{Fan}, X., {Strauss}, M.~A., {Becker}, R.~H., \& {et al.} 2006, AJ, 132, 117

\bibitem[{{Geha} {et~al.}(2003){Geha}, {Alcock}, {Allsman}, \& {et
  al.}}]{Geha2003AJ}
{Geha}, M., {Alcock}, C., {Allsman}, R.~A., \& {et al.} 2003, AJ, 125, 1

\bibitem[{{Giveon} {et~al.}(1999){Giveon}, {Maoz}, {Kaspi}, \& {et
  al.}}]{Giveon1999MNRAS}
{Giveon}, U., {Maoz}, D., {Kaspi}, S., \& {et al.} 1999, MNRAS, 306, 637

\bibitem[{{Green} {et~al.}(2004){Green}, {Silverman}, {Cameron}, \& {et
  al.}}]{Green2004ApJS}
{Green}, P.~J., {Silverman}, J.~D., {Cameron}, R.~A., \& {et al.} 2004, ApJS,
  150, 43

\bibitem[{{Hartman} {et~al.}(2008){Hartman}, {Gaudi}, {Holman}, \& {et
  al.}}]{Hartman2008ApJ}
{Hartman}, J.~D., {Gaudi}, B.~S., {Holman}, M.~J., \& {et al.} 2008, ApJ, 675,
  1254

\bibitem[{{Hawkins}(1993)}]{Hawkins1993Nature}
{Hawkins}, M.~R.~S. 1993, Nature, 366, 242

\bibitem[{{Hawkins}(2002)}]{Hawkins2002MNRAS}
---. 2002, MNRAS, 329, 76

\bibitem[{{Hook} {et~al.}(1994){Hook}, {McMahon}, {Boyle}, \& {et
  al.}}]{Hook1994MNRAS}
{Hook}, I.~M., {McMahon}, R.~G., {Boyle}, B.~J., \& {et al.} 1994, MNRAS, 268,
  305

\bibitem[{{Hornschemeier} {et~al.}(2001){Hornschemeier}, {Brandt}, {Garmire},
  \& {et al.}}]{Hornschemeier2001ApJ}
{Hornschemeier}, A.~E., {Brandt}, W.~N., {Garmire}, G.~P., \& {et al.} 2001,
  ApJ, 554, 742

\bibitem[{Hsu {et~al.}(2003)Hsu, Chang, \& Lin}]{libsvmguide}
Hsu, C.-W., Chang, C.-C., \& Lin, C.-J. 2003, A practical guide to support
  vector classification, Tech. rep., Department of Computer Science, National
  Taiwan University

\bibitem[{{Huertas-Company} {et~al.}(2008){Huertas-Company}, {Rouan}, {Tasca},
  \& {et al.}}]{Huertas2008AA}
{Huertas-Company}, M., {Rouan}, D., {Tasca}, L., \& {et al.} 2008, A\&A, 478,
  971

\bibitem[{{Ivezic} {et~al.}(2008){Ivezic}, {Tyson}, {Allsman}, {Andrew},
  {Angel}, \& {for the LSST Collaboration}}]{Ivezic2008arXiv}
{Ivezic}, Z., {Tyson}, J.~A., {Allsman}, R., {Andrew}, J., {Angel}, R., \& {for
  the LSST Collaboration}. 2008, ArXiv e-prints

\bibitem[{{Kaiser}(2004)}]{Kaiser2004SPIE}
{Kaiser}, N. 2004, in Society of Photo-Optical Instrumentation Engineers (SPIE)
  Conference Series, Vol. 5489, Society of Photo-Optical Instrumentation
  Engineers (SPIE) Conference Series, ed. {J.~M.~Oschmann Jr.}, 11--22

\bibitem[{{Kalfountzou} {et~al.}(2010){Kalfountzou}, {Trichas},
  {Rowan-Robinson}, \& {et al.}}]{Kalfountzou2010arXiv}
{Kalfountzou}, E., {Trichas}, M., {Rowan-Robinson}, M., \& {et al.} 2010, ArXiv
  e-prints

\bibitem[{{Kawaguchi} {et~al.}(1998){Kawaguchi}, {Mineshige}, {Umemura}, \& {et
  al.}}]{Kawaguchi1998ApJ}
{Kawaguchi}, T., {Mineshige}, S., {Umemura}, M., \& {et al.} 1998, ApJ, 504,
  671

\bibitem[{{Keller} {et~al.}(2002){Keller}, {Bessell}, {Cook}, \& {et
  al.}}]{Keller2002AJ}
{Keller}, S.~C., {Bessell}, M.~S., {Cook}, K.~H., \& {et al.} 2002, AJ, 124,
  2039

\bibitem[{{Kelly} {et~al.}(2009){Kelly}, {Bechtold}, \&
  {Siemiginowska}}]{Kelly2009ApJ}
{Kelly}, B.~C., {Bechtold}, J., \& {Siemiginowska}, A. 2009, ApJ, 698, 895

\bibitem[{{Kollmeier} {et~al.}(2006){Kollmeier}, {Onken}, {Kochanek}, \& {et
  al.}}]{Kollmeier2006ApJ}
{Kollmeier}, J.~A., {Onken}, C.~A., {Kochanek}, C.~S., \& {et al.} 2006, ApJ,
  648, 128

\bibitem[{{Koz{\l}owski} \& {Kochanek}(2009)}]{Kozlowski2009ApJ}
{Koz{\l}owski}, S., \& {Kochanek}, C.~S. 2009, ApJ, 701, 508

\bibitem[{{Koz{\l}owski} {et~al.}(2010){Koz{\l}owski}, {Kochanek}, {Udalski},
  \& {et al.}}]{Kozlowski2010ApJ}
{Koz{\l}owski}, S., {Kochanek}, C.~S., {Udalski}, A., \& {et al.} 2010, ApJ,
  708, 927

\bibitem[{{Kunder} \& {Chaboyer}(2008)}]{Kunder2008AJ}
{Kunder}, A., \& {Chaboyer}, B. 2008, AJ, 136, 2441

\bibitem[{{Lacy} {et~al.}(2004){Lacy}, {Storrie-Lombardi}, {Sajina}, \& {et
  al.}}]{Lacy2004ApJS}
{Lacy}, M., {Storrie-Lombardi}, L.~J., {Sajina}, A., \& {et al.} 2004, ApJS,
  154, 166

\bibitem[{{Laurent} {et~al.}(2000){Laurent}, {Mirabel}, {Charmandaris},
  {Gallais}, {Madden}, {Sauvage}, {Vigroux}, \& {Cesarsky}}]{Laurent2000AA}
{Laurent}, O., {Mirabel}, I.~F., {Charmandaris}, V., {Gallais}, P., {Madden},
  S.~C., {Sauvage}, M., {Vigroux}, L., \& {Cesarsky}, C. 2000, A\&A, 359, 887

\bibitem[{{Leinert} {et~al.}(2004){Leinert}, {van Boekel}, {Waters}, \& {et
  al.}}]{Leinert2004AA}
{Leinert}, C., {van Boekel}, R., {Waters}, L.~B.~F.~M., \& {et al.} 2004, A\&A,
  423, 537

\bibitem[{{Li} {et~al.}(2003){Li}, {Liu}, \& {Xie}}]{LiICCIMA2003}
{Li}, C., {Liu}, F., \& {Xie}, Y. 2003, Computational Intelligence and
  Multimedia Applications, International Conference on, 0, 37

\bibitem[{{Lomb}(1976)}]{Lomb1976ApSS}
{Lomb}, N.~R. 1976, Ap\&SS, 39, 447

\bibitem[{{Lupton} {et~al.}(2005){Lupton}, {Juri{\'c}}, {Ivezi{\'c}}, \& {et
  al.}}]{Lupton2005AAS}
{Lupton}, R.~H., {Juri{\'c}}, M., {Ivezi{\'c}}, Z., \& {et al.} 2005, in
  Bulletin of the American Astronomical Society, Vol.~37, Bulletin of the
  American Astronomical Society, 1384--+

\bibitem[{{MacLeod} {et~al.}(2010){MacLeod}, {Brooks}, {Ivezic}, \& {et
  al.}}]{MacLeod2010}
{MacLeod}, C.~L., {Brooks}, K., {Ivezic}, Z., \& {et al.} 2010, ArXiv e-prints

\bibitem[{{Malfait} {et~al.}(1998){Malfait}, {Bogaert}, \&
  {Waelkens}}]{Malfait1998AA}
{Malfait}, K., {Bogaert}, E., \& {Waelkens}, C. 1998, A\&A, 331, 211

\bibitem[{{Meixner} {et~al.}(2006){Meixner}, {Gordon}, {Indebetouw}, \& {et
  al.}}]{Meixner2006AJ}
{Meixner}, M., {Gordon}, K.~D., {Indebetouw}, R., \& {et al.} 2006, AJ, 132,
  2268

\bibitem[{{Mennickent} {et~al.}(2002){Mennickent}, {Pietrzy{\'n}ski}, {Gieren},
  \& {et al.}}]{Mennickent2002AA}
{Mennickent}, R.~E., {Pietrzy{\'n}ski}, G., {Gieren}, W., \& {et al.} 2002,
  A\&A, 393, 887

\bibitem[{{Metcalf} \& {Madau}(2001)}]{Metcalf2001ApJ}
{Metcalf}, R.~B., \& {Madau}, P. 2001, ApJ, 563, 9

\bibitem[{{Miranda} \& {Macci{\`o}}(2007)}]{Miranda2007MNRAS}
{Miranda}, M., \& {Macci{\`o}}, A.~V. 2007, MNRAS, 382, 1225

\bibitem[{{Nilsson} {et~al.}(2006){Nilsson}, J., {Bj{\"o}rkegren}, \& {et
  al.}}]{Nilsson2006springer}
{Nilsson}, R., J., P., {Bj{\"o}rkegren}, J., \& {et al.} 2006, Lecture Notes in
  Computer Science, Vol. 4212, Evaluating Feature Selection for SVMs in High
  Dimensions, ed. J.~F{\"u}rnkranz, T.~Scheffer, \& M.~Spiliopoulou (Springer
  Berlin, Heidelberg), 719--726

\bibitem[{{Panik}(2005)}]{Panik2005}
{Panik}, M.~J. 2005, Advanced statistics from an elementary point of view (San
  Diego, CA: Elsevier Academic Press), 576

\bibitem[{{Peng} {et~al.}(2006){Peng}, {Impey}, {Rix}, \& {et
  al.}}]{Peng2006ApJ}
{Peng}, C.~Y., {Impey}, C.~D., {Rix}, H., \& {et al.} 2006, ApJ, 649, 616

\bibitem[{{Platt}(1999)}]{Platt1999}
{Platt}, J.~C. 1999, in Advances in Large Margin Classifiers (MIT Press),
  61--74

\bibitem[{{Press}(1969)}]{Press1969}
{Press}, S.~J. 1969, The Annals of Mathematical Statistics, 40, 188

\bibitem[{{Press} \& {Rybicki}(1989)}]{Press1989ApJ}
{Press}, W.~H., \& {Rybicki}, G.~B. 1989, ApJ, 338, 277

\bibitem[{{Press} {et~al.}(1992){Press}, {Teukolsky}, {Vetterling}, \& {et
  al.}}]{Press1992}
{Press}, W.~H., {Teukolsky}, S.~A., {Vetterling}, W.~T., \& {et al.} 1992,
  {Numerical recipes in C. The art of scientific computing}, ed. {Press, W.~H.,
  Teukolsky, S.~A., Vetterling, W.~T., \& Flannery, B.~P. }

\bibitem[{{Rees}(1984)}]{Rees1984ARAA}
{Rees}, M.~J. 1984, ARA\&A, 22, 471

\bibitem[{{Reeves} \& {Turner}(2000)}]{Reeves2000MNRAS}
{Reeves}, J.~N., \& {Turner}, M.~J.~L. 2000, MNRAS, 316, 234

\bibitem[{{Richards} {et~al.}(2006){Richards}, {Strauss}, {Fan}, \& {et
  al.}}]{Richards2006AJ}
{Richards}, G.~T., {Strauss}, M.~A., {Fan}, X., \& {et al.} 2006, AJ, 131, 2766

\bibitem[{{Ross} {et~al.}(2009){Ross}, {Shen}, {Strauss}, \& {et
  al.}}]{Ross2009ApJ}
{Ross}, N.~P., {Shen}, Y., {Strauss}, M.~A., \& {et al.} 2009, ApJ, 697, 1634

\bibitem[{{Scargle}(1982)}]{Scargle1982ApJ}
{Scargle}, J.~D. 1982, ApJ, 263, 835

\bibitem[{{Schild} {et~al.}(2009){Schild}, {Lovegrove}, \&
  {Protopapas}}]{Schild2009AJ}
{Schild}, R.~E., {Lovegrove}, J., \& {Protopapas}, P. 2009, AJ, 138, 421

\bibitem[{{Schmidt} {et~al.}(2010){Schmidt}, {Marshall}, {Rix}, \& {et
  al.}}]{Schmidt2010ApJ}
{Schmidt}, K.~B., {Marshall}, P.~J., {Rix}, H., \& {et al.} 2010, ApJ, 714,
  1194

\bibitem[{{Schmidtke} {et~al.}(1999){Schmidtke}, {Cowley}, {Crane}, \& {et
  al.}}]{Schmidtke1999AJ}
{Schmidtke}, P.~C., {Cowley}, A.~P., {Crane}, J.~D., \& {et al.} 1999, AJ, 117,
  927

\bibitem[{{Sesar} {et~al.}(2007){Sesar}, {Ivezi{\'c}}, {Lupton}, \& {et
  al.}}]{Sesar2007AJ}
{Sesar}, B., {Ivezi{\'c}}, {\v Z}., {Lupton}, R.~H., \& {et al.} 2007, AJ, 134,
  2236

\bibitem[{{Shen} {et~al.}(2007){Shen}, {Strauss}, {Oguri}, \& {et
  al.}}]{Shen2007AJ}
{Shen}, Y., {Strauss}, M.~A., {Oguri}, M., \& {et al.} 2007, AJ, 133, 2222

\bibitem[{{Shin} {et~al.}(2009){Shin}, {Sekora}, \& {Byun}}]{Shin2009MNRAS}
{Shin}, M., {Sekora}, M., \& {Byun}, Y. 2009, MNRAS, 400, 1897

\bibitem[{{Simcoe} {et~al.}(2004){Simcoe}, {Sargent}, \&
  {Rauch}}]{Simcoe2004ApJ}
{Simcoe}, R.~A., {Sargent}, W.~L.~W., \& {Rauch}, M. 2004, ApJ, 606, 92

\bibitem[{{Singh} {et~al.}(1995){Singh}, {Drake}, \& {White}}]{Singh1995ApJ}
{Singh}, K.~P., {Drake}, S.~A., \& {White}, N.~E. 1995, ApJ, 445, 840

\bibitem[{{Stern} {et~al.}(2005){Stern}, {Eisenhardt}, {Gorjian}, \& {et
  al.}}]{Stern2005ApJ}
{Stern}, D., {Eisenhardt}, P., {Gorjian}, V., \& {et al.} 2005, ApJ, 631, 163

\bibitem[{{Stetson}(1996)}]{Stetson1996PASP}
{Stetson}, P.~B. 1996, PASP, 108, 851

\bibitem[{{Stetson}(2000)}]{Stetson2000PASP}
---. 2000, PASP, 112, 925

\bibitem[{{Sumi} {et~al.}(2005){Sumi}, {Wo{\'z}niak}, {Eyer}, \& {et
  al.}}]{Sumi2005MNRAS}
{Sumi}, T., {Wo{\'z}niak}, P.~R., {Eyer}, L., \& {et al.} 2005, MNRAS, 356, 331

\bibitem[{{Terlevich} {et~al.}(1992){Terlevich}, {Tenorio-Tagle}, {Franco}, \&
  {et al.}}]{Terlevich1992MNRAS}
{Terlevich}, R., {Tenorio-Tagle}, G., {Franco}, J., \& {et al.} 1992, MNRAS,
  255, 713

\bibitem[{{Thomas} {et~al.}(2005){Thomas}, {Griest}, {Popowski}, \& {et
  al.}}]{Thomas2005ApJ}
{Thomas}, C.~L., {Griest}, K., {Popowski}, P., \& {et al.} 2005, ApJ, 631, 906

\bibitem[{{Trichas} {et~al.}(2010){Trichas}, {Rowan-Robinson}, {Georgakakis},
  \& {et al.}}]{Trichas2010MNRAS}
{Trichas}, M., {Rowan-Robinson}, M., {Georgakakis}, A., \& {et al.} 2010,
  MNRAS, 405, 2243

\bibitem[{{Tsalmantza} {et~al.}(2007){Tsalmantza}, {Kontizas}, {Bailer-Jones},
  \& {et al.}}]{Tsalmantza2007AA}
{Tsalmantza}, P., {Kontizas}, M., {Bailer-Jones}, C.~A.~L., \& {et al.} 2007,
  A\&A, 470, 761

\bibitem[{{Udalski} {et~al.}(1997){Udalski}, {Kubiak}, \&
  {Szymanski}}]{Udalski1997AcA}
{Udalski}, A., {Kubiak}, M., \& {Szymanski}, M. 1997, Acta Astronomica, 47, 319

\bibitem[{{Udalski} {et~al.}(2008){Udalski}, {Szymanski}, {Soszynski}, \& {et
  al.}}]{Udalski2008AcA}
{Udalski}, A., {Szymanski}, M.~K., {Soszynski}, I., \& {et al.} 2008, Acta
  Astronomica, 58, 69

\bibitem[{{Vanden Berk} {et~al.}(2004){Vanden Berk}, {Wilhite}, {Kron}, \& {et
  al.}}]{VandenBerk2004ApJ}
{Vanden Berk}, D.~E., {Wilhite}, B.~C., {Kron}, R.~G., \& {et al.} 2004, ApJ,
  601, 692

\bibitem[{{Viel} {et~al.}(2002){Viel}, {Matarrese}, {Mo}, \& {et
  al.}}]{Viel2002MNRAS}
{Viel}, M., {Matarrese}, S., {Mo}, H.~J., \& {et al.} 2002, MNRAS, 329, 848

\bibitem[{{von Neumann}(1941)}]{Neumann1941}
{von Neumann}, J. 1941, Ann. Math. Statist., 12, 367

\bibitem[{{Wachman} {et~al.}(2009){Wachman}, {Khardon}, {Protopapas}, \& {et
  al.}}]{Wachman2009}
{Wachman}, G., {Khardon}, R., {Protopapas}, P., \& {et al.} 2009, in ECML PKDD
  '09: Proceedings of the European Conference on Machine Learning and Knowledge
  Discovery in Databases (Berlin, Heidelberg: Springer-Verlag), 489--505

\bibitem[{{Wadadekar}(2005)}]{Wadadekar2005PASP}
{Wadadekar}, Y. 2005, PASP, 117, 79

\bibitem[{{Watson} {et~al.}(2009){Watson}, {Schr{\"o}der}, {Fyfe}, \& {et
  al.}}]{Watson2009AA}
{Watson}, M.~G., {Schr{\"o}der}, A.~C., {Fyfe}, D., \& {et al.} 2009, A\&A,
  493, 339

\bibitem[{{Weston} {et~al.}(2001){Weston}, {Mukherjee}, {Chapelle}, \& {et
  al.}}]{Weston2001NIPS}
{Weston}, J., {Mukherjee}, S., {Chapelle}, O., \& {et al.} 2001, in Advances in
  Neural Information Processing Systems 13, Vol.~13 (MIT Press), 668--674

\bibitem[{{Wonnacott} {et~al.}(1994){Wonnacott}, {Kellett}, {Smalley}, \&
  et~al.}]{Wonnacott1994MNRAS}
{Wonnacott}, D., {Kellett}, B.~J., {Smalley}, B., \& et~al. 1994, MNRAS, 267,
  1045

\bibitem[{{Wood}(2000)}]{Wood2000PASA}
{Wood}, P.~R. 2000, Publications of the Astronomical Society of Australia, 17,
  18

\bibitem[{{Wo{\'z}niak}(2000)}]{Wozniak2000AcA}
{Wo{\'z}niak}, P.~R. 2000, Acta Astronomica, 50, 421

\bibitem[{{Wozniak} {et~al.}(2002){Wozniak}, {Udalski}, {Szymanski}, \& {et
  al.}}]{Wozniak2002AcA}
{Wozniak}, P.~R., {Udalski}, A., {Szymanski}, M., \& {et al.} 2002, Acta
  Astronomica, 52, 129

\bibitem[{{Wo{\'z}niak} {et~al.}(2004{\natexlab{a}}){Wo{\'z}niak}, {Vestrand},
  {Akerlof}, \& {et al.}}]{Wozniak2004AJ}
{Wo{\'z}niak}, P.~R., {Vestrand}, W.~T., {Akerlof}, C.~W., \& {et al.}
  2004{\natexlab{a}}, AJ, 127, 2436

\bibitem[{{Wo{\'z}niak} {et~al.}(2004{\natexlab{b}}){Wo{\'z}niak}, {Williams},
  {Vestrand}, \& {et al.}}]{Wozniak2004AJb}
{Wo{\'z}niak}, P.~R., {Williams}, S.~J., {Vestrand}, W.~T., \& {et al.}
  2004{\natexlab{b}}, AJ, 128, 2965

\bibitem[{{Zackrisson} {et~al.}(2003){Zackrisson}, {Bergvall}, {Marquart}, \&
  {et al.}}]{Zackrisson2003AA}
{Zackrisson}, E., {Bergvall}, N., {Marquart}, T., \& {et al.} 2003, A\&A, 408,
  17

\bibitem[{{Zebrun} {et~al.}(2001){Zebrun}, {Soszynski}, {Wozniak}, \& {et
  al.}}]{Zebrun2001AcA}
{Zebrun}, K., {Soszynski}, I., {Wozniak}, P.~R., \& {et al.} 2001, Acta
  Astronomica, 51, 317

\bibitem[{{Zhang} \& {Zhao}(2004)}]{Zhang2004AA}
{Zhang}, Y., \& {Zhao}, Y. 2004, A\&A, 422, 1113

\end{thebibliography}

\section*{Appendix}
\label{sec:appendix}

In this appendix, we introduce the 11 time series features 
including four new features that we have developed for this work
and the remaining  seven features.\\

{\underline{Four new time series features}}:
\begin{itemize}
\label{sec:Features}

\item{Three autocorrelation Indices}
\label{sec:autocorrelation_indices}

These three indices are based on the autocorrelation function.
The autocorrelation function is defined as:

\begin{eqnarray}
AC(\tau) = \displaystyle\frac{1}{(N - \tau) \,\, \sigma^{2}} \, \sum_{i=1}^{N-\tau}(m_{i} - \bar{m})(m_{i + \tau} - \bar{m}),
\end{eqnarray}

{\noindent}where 
$N$ is the total number of data points,
$\tau = 1, 2, \ldots, N-1 $ is the time lag,
$\sigma$ is the standard deviation,
$m$ is the magnitude,
$i$ is the index for each data point and
$\bar{m}$ is the mean magnitude.
Figure \ref{fig:AC} shows the $AC(\tau)$ for various types of variables and non-variables 
extracted from the MACHO database.
Note that, in each panel, we show the $AC(\tau)$ of multiple objects of that type
to demonstrate the overall $AC(\tau)$ patterns.
We used more than 50 objects of non-variables, RR lyraes, Cepheids,
eclipsing binaries and microlensing events. The overall $AC(\tau)$ patterns
were preserved even if we used more objects (i.e. several hundreds).
For long period variables (LPVs), Be stars and QSOs, we used about 10 object of each type to
show individual $AC(\tau)$ pattern.
The x-axis is the time lag, $\tau$  in days and the y-axis is the autocorrelation value.
As the figure shows, non-variables and all periodic variables but LPVs
show different $AC(\tau)$ patterns from QSOs, Be stars, LPVs and microlensing events.
\citet{Schild2009AJ} also noticed that the  $AC(\tau)$
could be useful for discovering QSOs.
Thus, by quantifying the $AC(\tau)$, we can separate certain types of variables.
In the following paragraphs, we introduce three 
time series features that we are using to quantify $AC(\tau)$.

\begin{figure*}[tbp]
\begin{center}
       \includegraphics[width=1.0\textwidth]{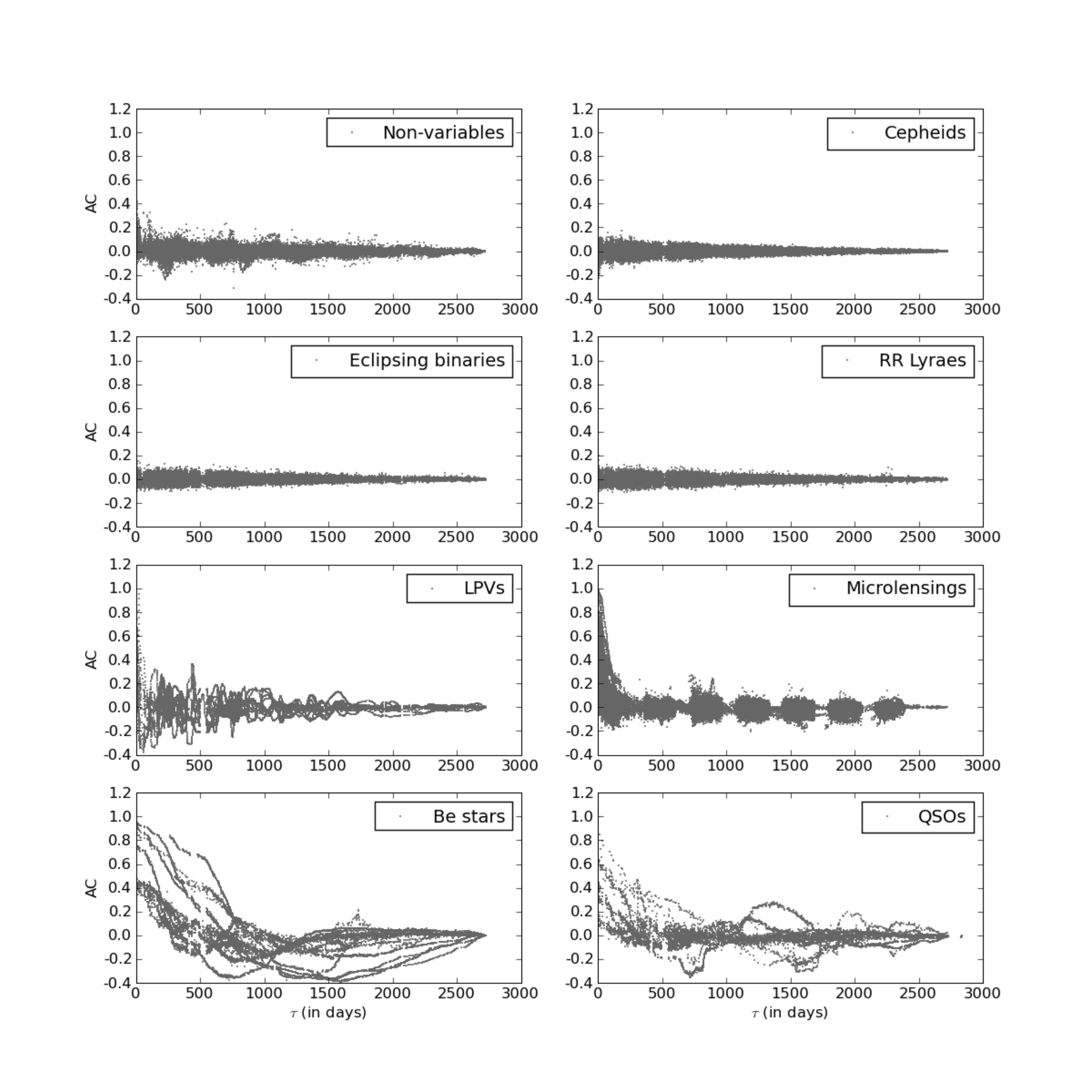}
\end{center}
    \caption{Set of autocorrelation functions of variable and non-variable stars.
    The x-axis is the time lag, $\tau$, in days, and the y-axis is the autocorrelation function value.
	Non-variable stars, Cepheids, eclipsing binaries and RR Lyraes show
    different patterns from QSOs, Be stars, LPVs and microlensing events.}
    \label{fig:AC}
\end{figure*}

\begin{itemize}

\item{$N_{above}$, $N_{below}$ }

We constructed empirical boundary lines on the  AC vs $\tau$ diagram  to 
separate non-variables and periodic variables from others.
To do so, we calculated  the average and standard deviation  of the autocorrelation functions 
for  non-variables and periodic variables (except LPVs), for each time lag $\tau$.
We then constructed upper and lower boundary lines to be $\pm4 \sigma$  from the average line.
Figure \ref{fig:AC_boundary} shows the calculated upper and lower boundary lines\footnote{
We removed fluctuated data points using moving average.}.
To derive $N_{above}$ and $N_{below}$  for each lightcurve, 
we counted the number of points above, $N_{above}$, and number of points below, $N_{below}$, these lines.

\item{Stetson $K$}

Stetson $K$ (Eq. \ref{eq:StetsonK}) was defined to observe 
the distribution of measurements between
the maximum and minimum values of the measurements \citep{Stetson1996PASP}.
For details including the definition of Stetson $K$, see the Appendix.
We used Stetson $K$ to characterize the different $AC(\tau)$ patterns, Stetson $K_{AC}$.

\end{itemize}

\item{$R_{cs}$}

$R_{cs}$ is the range of a cumulative sum \citep{Ellaway1978} of each lightcurve and is defined as:

\begin{eqnarray}
\begin{array}{l}
R_{cs} = {\rm{max}}(S) - {\rm{min}}(S),
\\ \\
S_{l} = \displaystyle \frac{1}{N \,\, \sigma} \sum_{i=1}^{l}(m_{i} - \bar{m}) \, ,
\end{array}
\end{eqnarray}

{\noindent}where max (min) is the maximum (minimum) value of $S$ and $l=1, 2, \ldots, N$.
$R_{cs}$ is typically large for LPVs, microlensing events, Be stars and QSOs
while it is relatively small for non-variables and other periodic variables such as 
RR Lyraes, Cepheids and eclipsing binaries.

\begin{figure}[tbp]
\begin{center}
       \includegraphics[width=0.45\textwidth]{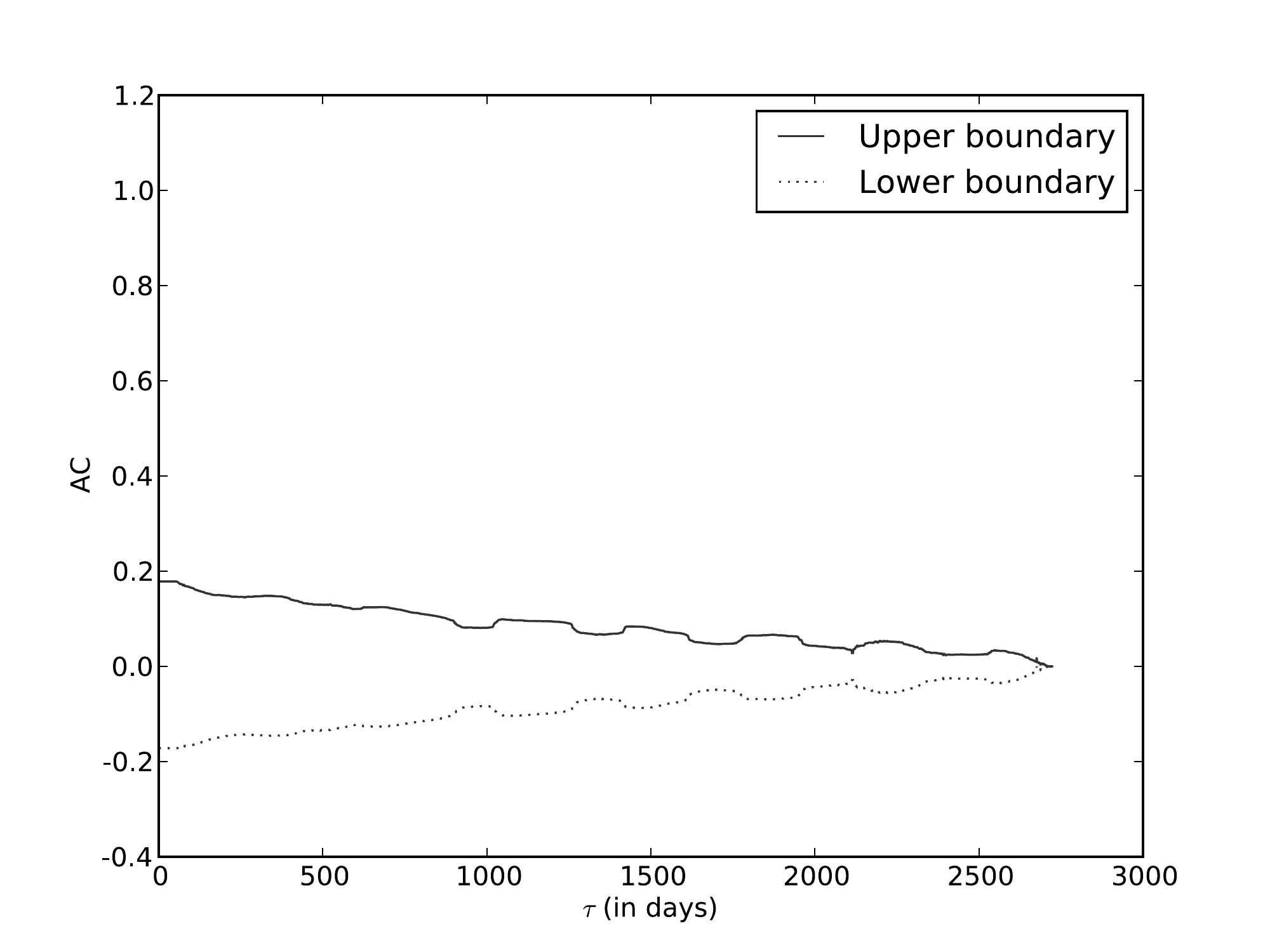}
\end{center}
    \caption{Two boundary lines constructed using autocorrelation functions 
    of non-variable stars, eclipsing binaries, RR Lyraes and Cepheids. 
    The x-axis is the time lag, $\tau$, in days, and the y-axis is the autocorrelation value.
    Based on the    lines, we derived $N_{above}$ and $N_{below}$. See the text for details.
    }
    \label{fig:AC_boundary}
\end{figure}

\end{itemize}

{\underline{Other seven time series features}}:
\begin{itemize}

\item{$\displaystyle\frac{\sigma}{\bar{m}}$} 

This is a simple variability index and is
defined as the ratio of the standard deviation, $\sigma$, to the mean magnitude, $\bar{m}$.
If a lightcurve has strong variability, $\sigma/\bar{m}$ of the lightcurve is generally large.

\item{Period and Period S/N}

To derive periods and their signal to noise ratios (S/N), 
we employed the Lomb-Scargle algorithm \citep{Lomb1976ApSS, Scargle1982ApJ, Press1989ApJ, Press1992}.
We search for periods between 0.1 and 1000 days\footnote{We used 
\href{http://www.cfa.harvard.edu/~jhartman/vartools/}{\fontfamily{pcr}\selectfont VARTOOLS} \citep{Hartman2008ApJ} for deriving 
periods and period S/Ns.},
which covers not only short period variable stars such as RR Lyraes, 
Cepheids and eclipsing binaries but also LPVs.
Among the detected periods, we selected the period with the highest S/N.
The S/N of each period is calculated based on the Lomb periodogram \citep{Scargle1982ApJ, Press1992}.

\item{Stetson $L$}
\label{item:stetson_L}

Stetson $L$ variability index \citep{Stetson1996PASP} describes
the synchronous variability of different bands and is defined as:

\begin{eqnarray}
L = \frac{JK}{0.798} \, ,
\end{eqnarray} 

{\noindent}where $J$ and $K$ are different Stetson indices.
Stetson $J$ is calculated based on two simultaneous lightcurves of a same star 
(e.g. MACHO B and R bands) and is defined as:

\begin{eqnarray}
\begin{array}{l}
\displaystyle
J = \frac{1}{N}\sum_{i=1}^{N}{\rm{sgn}}(P_{i})\sqrt{|P_{i}|},
\\ \\
\displaystyle
P_{i} = \delta_{p}(i)\, \delta_{q}(i),
\\ \\
\displaystyle
\delta_{p}(i) = \sqrt{\frac{N}{N-1}} \, \, \frac{m_{p,i} - \bar{m}}{\sigma_{p,i}},
\end{array}
\end{eqnarray}

{\noindent}where $i$ is the index for each data point, $N$ is the total number of data points, 
sgn($P_{i}$) is the sign of $P_{i}$ and $m$ is the magnitude.
$p$ and $q$ indicate two different bands. 
$\sigma_{p,i}$ is the standard error of $i_{th}$ magnitude of band $p$.
In case of the MACHO database, 
$p$ and $q$ indicate the MACHO B and R bands.
To derive $J$ from each MACHO time series, 
we used only the data points which have observations from both MACHO B and R bands at the same epoch.

Steston $K$ is calculated using a single band lightcurve and is defined as:

\begin{eqnarray}
K = \frac{1}{\sqrt{N}} \frac{\sum_{i=1}^{N}|\delta(i)|} {\sqrt{\sum_{i=1}^{N}\delta(i)^{2}}}.
\label{eq:StetsonK}
\end{eqnarray}

It is known that $K = 0.900$ for a pure sinusoid and 0.798 for a Gaussian distribution. 
For details, see \citet{Stetson1996PASP}.

In brief, Stetson $L$ is generally large for achromatic variable sources 
and small for non-variables or chromatic variables.

\item{$\eta$}
\label{item:eta}

Variability index $\eta$ is the ratio of the mean of the square of successive differences to the variance of data points.
The index was originally proposed to check whether the successive data points are independent or not.
In other words, the index was developed to check if any trends exist in the data \citep{Neumann1941}. It is 
defined as:

\begin{equation}
\eta = \frac{1}{(N-1) \,\, \sigma^{2}} \sum_{i=1}^{N-1}(m_{i+1} - m_{i})^{2} .
\end{equation}

The index has been substantially investigated by several authors (see \citealt{Neumann1941, Press1969} and references therein).
In brief, if there exists positive  serial correlation, $\eta$ is relatively small.
On the other hand, if there exists negative serial correlation, $\eta$ is large.
\citet{Shin2009MNRAS} used $\eta$ to select variable candidates
from the Northern Sky Variability Survey database \citep{Wozniak2004AJ}.

As the top right panel of Figure \ref{fig:index_index} shows, $\eta$ is relatively small for 
the variables which have positive autocorrelation such as QSOs, Be stars and LPVs.
Non-variables or microlensing events show large $\eta$ 
since they do not have strong positive correlation.
In the cases of other periodic variables such as RR Lyraes, Cepheids and eclipsing binaries,
$\eta$ is also relatively large even though they are periodic variables and 
therefore have positive correlation.
This is because 1) we derive $\eta$ not from the folded MACHO lightcurves but from the original lightcurves
and 2) MACHO observed a field a few times  per week,
which is not enough to reveal positive correlation for small time scales.
In other words, most raw MACHO lightcurves of the periodic variables 
do not have strong positive correlation and thus have large $\eta$.

\item{$B-R$}

We used an average color for each MACHO lightcurve as:

\begin{equation}
B-R = \displaystyle\bar{m}_{{B}_{M}} - \bar{m}_{{R}_{M}},
\end{equation}

{\noindent}where $\bar{m}_{{B}_{M}}$, $\bar{m}_{{R}_{M}}$ are the mean magnitudes of MACHO B, R bands.

Color information, $B-R$, is useful in separating QSOs from 
some other types of variables as several other studies suggested 
\citep{Giveon1999MNRAS, Eyer2002AcA, Geha2003AJ}.
Nevertheless it is known that color\footnote{SDSS $u-g$ color. See \citealt{Schmidt2010ApJ} for details.}
is not very efficient discriminator for
selecting intermediate redshift QSOs (i.e. $2.5 < z < 3.0$) 
although it is efficient for selecting high and low redshift QSOs
\citep{Richards2006AJ, Schmidt2010ApJ}.
Note that we used not only color information but also 
other multiple time series features derived solely based 
on the variability characteristics of lightcurves,
which helps to identify intermediate redshift QSOs as well as
high and low redshift QSOs.

\item{$Con$}

The index was introduced for the selection of variable stars
from the OGLE database \citep{Wozniak2000AcA}. 
To calculate $Con$, we counted the number of three consecutive data points that
are brighter or fainter than 2$\sigma$ and normalized the number by $N-2$.
$Con$ is close to zero for non-variable stars while it is relatively large for variables.
In addition, $Con$ is relatively large for the long-time scale varying sources such as LPVs
because such variables tend to have plenty of consecutive data points bigger than 2$\sigma$.

\end{itemize}

\end{document}